\documentclass[%
 reprint,
 amsmath,amssymb,
 aps,
prb,
]{revtex4-2}

\usepackage{graphicx}
\usepackage{dcolumn}
\usepackage{bm}
\usepackage{color}

\begin{document}

\title{Stability, growth, and doping of In$_{2}$(Si, Ge)$_{2}$O$_{7}$ as promising n-type wide-gap semiconductors}

\author{Cheng-Wei Lee}

\author{Kingsley Egbo}

\author{Emily Garrity}

\author{Matthew Jankousky}

\author{Henry Garland}

\author{Andriy Zakutayev}
\email{andriy.zakutayev@nrel.gov}

\author{Vladan Stevanovi\'c}
\email{vstevano@mines.edu}
\affiliation{Colorado School of Mines, Golden, CO 80401,USA}
\affiliation{National Renewable Energy Laboratory, Golden, CO 80401,USA }


\begin{abstract}
In this paper we investigate, computationally and experimentally, the phase stability, electronic structure properties, and the propensity for n-type doping of In$_{2}$X$_{2}$O$_{7}$ (X=Si, Ge) ternary oxides.  This family of materials contains promising novel wide-gap semiconductors based on their estimated high \textit{n}-type Baliga figures of merit and acceptable thermal conductivity for power electronics applications. Here, we find that both In$_{2}$Si$_{2}$O$_{7}$ and In$_{2}$Ge$_{2}$O$_{7}$ to be n-type dopable, with Zr providing between 10$^{16}$ and above 10$^{21}$ cm$^{-3}$ net donor concentrations under O-poor conditions, depending on the chemistry, structure (ground-state thorvetite or high-pressure pyrochlore) and synthesis temperature. Initial thin-film growth and annealing leads to polycrystalline In$_{2}$Ge$_{2}$O$_{7}$ thin films in thorvetite structure with band gap over 4 eV, and confirms Zr doping predictions by achieving electron concentrations at 10$^{14}$-10$^{16}$ cm$^{-3}$ under O-rich condition. While future epitaxial growth development is still needed, this study establishes In$_{2}$X$_{2}$O$_{7}$ as promising n-type wide-gap semiconductors for power electronic applications. 
\end{abstract}

\maketitle


\section{Introduction}
%
The expected widespread electrification of transportation in combination with the increased adoption of renewable energy generation, will likely demand novel types of power electronic devices which can efficiently process high powers of 100 kWs to 10 MWs \cite{blaabjerg_power_2023,IEC_power_2023}. The power ratings of present devices, including AC-DC converters for photovoltaics, wind, and electric vehicle chargers, are largely limited by the performance of the active semiconductor material \cite{kaplar_generation-after-next_2017}.  One approach to enhance device performance is to engineer around the problems of existing commercial power semiconductors (Si, SiC, GaN). The alternative approach, which is discussed in this work, is to identify novel semiconductors with inherently better properties, including ultra wide band gap (UWBG) oxide semiconductors.\cite{Garrity_PRXEnergy_2022,gorai_computational_2019} An additional advantage of the oxides is that compared to carbides and nitrides, using oxides removes concerns of oxidation and other degradation at higher temperatures. Also, their lower melting points compared to carbides and nitrides hold promise for less demanding synthesis and processing techniques, as exemplified by the melt growth of $\beta$-Ga$_2$O$_3$ \cite{reese_how_2019}.

\begin{figure}
\includegraphics[width=\linewidth]{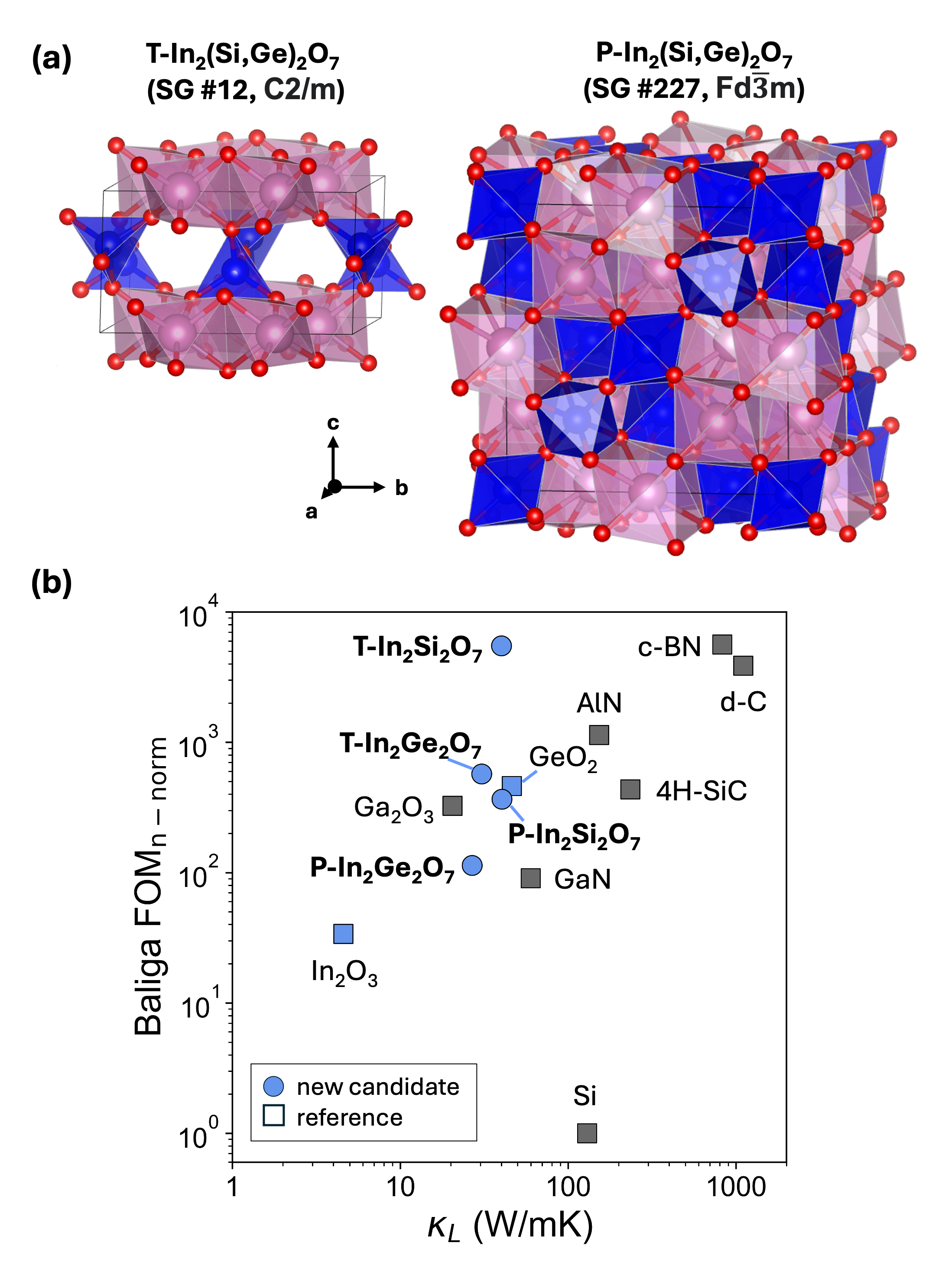}
\caption{\label{fig1} (a) Models of In$_{2}$X$_{2}$O$_{7}$ (X=Si,Ge - blue) in the ground-state thortveitite (T-) and high-pressure pyrochlore (P-) crystal structures. (b) Results from our previous high-throughput screening for promising novel power semiconductors \cite{Garrity_PRXEnergy_2022, gorai_computational_2019}. The Baliga figure of merit is calculated using electronic structures from the HSE06 method and is normalized relative to that of $n$-type doped crystalline Si. Points represent known power electronics semiconductors (grey squares), ``parent" binaries identified in other works (blue squares), and the In$_{2}$X$_{2}$O$_{7}$ investigated here (blue circles).}
\end{figure}

In our previous high-throughput computational materials searches \cite{Garrity_PRXEnergy_2022,gorai_computational_2019} we identified the In$_{2}$X$_{2}$O$_{7}$ (X=Si, Ge) group of oxides as particularly promising for \textit{n}-type vertical power devices, in addition to  II-IV spinels and calcite-type borates \cite{Garrily_CM_2023_IIIBO3}. This includes both the ground state thortveitite (space group 12, C2/m) and high-pressure pyrochlore (space group 227, Fd$\overline{3}$m) structures for both In$_2$Ge$_2$O$_7$ and In$_2$Si$_2$O$_7$. This assessment is based on their predicted bulk properties as described by the thermal conductivity and Baliga figure of merit (BFOM) \cite{baliga_semiconductors_1982}. A larger BFOM describes smaller on-state conduction losses in unipolar field-effect transistors and scales strongly with the band gap. The calculated \textit{n}-type BFOMs of the In$_{2}$X$_{2}$O$_{7}$ compounds span two orders of magnitude, from about 100 times greater than $n$-type crystalline Si to more than 5000 times greater. As evident from Fig.~\ref{fig1} this range extends above current state-of-the-art power semiconductors like GaN and 4H-SiC. While the predicted thermal conductivities are below that of SiC (and cubic BN or diamond) the potential improvement in BFOM represents a reduction in  power losses and associated heat and thus may reduce the need for thermal management. 

An additional reason to study In$_2$Ge$_2$O$_7$ is that this ternary compound is expected to form as an interfacial layer between the emerging GeO$_2$ UWBG oxide semiconductor\cite{Chae_APL_2019,Mengle_JAP_2019,Bushick_APL_2020} and the common In$_2$O$_3$ conductive oxide n-type contacts. Our previous calculations of phase equilibria between wide-gap semiconductors and contact materials identified  In$_2$Ge$_2$O$_7$ as the likely interphase at the  In$_2$O$_3$/GeO$_2$ interface.\cite{Lee_ACEELM_2024} However, very little is know about In$_2$Ge$_2$O$_7$  band structure, point defects, electrical doping and other physical properties, all of which would be important for such interfacial layers. In particular, the doping evaluation of emerging UWBG oxides is important because while the Baliga FOM assumes that these materials can be doped $n$-type to the desired charge carrier concentration (typically between 10$^{15}$ and 10$^{20}$ cm$^{-3}$) \cite{tsao_ultrawide-bandgap_2018}, realizing the necessary charge carrier concentrations in wide band gap materials has been historically difficult \cite{yan_doping_2008,Goyal_CM:2020}. In other materials systems, formation of thin interfacial layers, such as Ga$_2$NiO$_4$ at the Ga$_2$O$_3$/NiO interface,\cite{Egbo_APL_2024} has been shown to beneficially influence the power electronic device performance and reliability. 

In prior literature, powders of  In$_{2}$X$_{2}$O$_{7}$ (X=Si, Ge) in the thortveitite structure have been synthesized using solid-state reactions for both In$_2$Ge$_2$O$_7$ \cite{gaewdang_structural_1994,li_comparative_2015,shannon_synthesis_1970} and In$_2$Si$_2$O$_7$ \cite{gaewdang_structural_1994, reid_high-pressure_1977}. The pyrochlore structures are realizable using moderately high pressures and temperatures --- 1373-1573 K at 5.2-6.5 GPa for In$_2$Ge$_2$O$_7$ \cite{li_comparative_2015, shannon_synthesis_1968, messous_indium_1995} and 1273-1373 K at 7-12 GPa for In$_2$Si$_2$O$_7$ \cite{reid_high-pressure_1977, messous_indium_1995}. Additionally, single crystals of the thortveitite In$_2$Si$_2$O$_7$ measuring a few mm$^2$ and 1 mm thickness were grown using solvent flux growth \cite{messous_indium_1995}. This evidence of previous synthesis is promising for future growth of these materials.

In this paper we investigate properties of the In$_{2}$X$_{2}$O$_{7}$ (X=Si, Ge) group of materials experimentally and computationally, with the focus on stability, electronic structure, and electrical doping. Our study covers thermodynamic phase diagrams, polymorphic energies, electronic properties, defect chemistry, and doping behavior. We predict the oxygen partial pressure (pO$_2$) vs. temperature phase diagrams,  show that these compounds should be $n$-type dopable using Zr as an extrinsic dopant, synthesize and characterize Zr:In$_2$Ge$_2$O$_7$ polycrystalline thin films, and discuss the transition from the T- to the P- polymorphs. All of the predicted and measured properties strengthen our initial assessment of these In$_{2}$X$_{2}$O$_{7}$ (X=Si, Ge) compounds as potential novel UWBG oxide semiconductors for power electronic device applications.  

\section{Methods}\label{sec:methods}

\subsection{Stability and metastability}\label{sec:methods_stability}
%
The stability against decomposition of all four In$_2$(Si,Ge)$_2$O$_7$ phases was assessed using the well-established fitted elemental reference energies (FERE) method \cite{Stevanovic_PRB:2012} --- the ad-hoc corrections developed to fix the problems of density functional theory (DFT) and provide more accurate predictions of the compound enthalpies of formation. The values are tabulated on Table S1. Using FERE requires a specific numerical setup for the underlying DFT calculations, which was used throughout this work. In all DFT calculations the electron-electron interactions are treated using the Perdew-Burke-Ernzerhof (PBE)exchange-correlation functional \cite{PBE_PRL:1996} and the projector augmented wave (PAW) method \cite{PAW_PRB:1994} as implemented in the VASP computer code \cite{Kresse_PRB_1999}. Plane-wave cutoff of 340 eV is used in all calculations, which is at least $\sim$20 \% higher than the recommended value for all pseudopotentials. Automatic generation of the $\Gamma$-centered k-point grid is employed with the $R_k$ value of 20. Starting crystal structures for both chemistries were obtained from the Inorganic Crystal Structures Database (ICSD) \cite{icsd}. These were relaxed to the corresponding DFT local minima by relaxing all degrees of freedom, including volume,
cell shape and atomic positions. Volume and cell-shape relaxations are restarted at least three times for the real-space grid to be re-created. Structural relaxations are considered converged when maximal force on any atom dips below 0.02 eV/\AA, total energy stops changing by more than $10^{-5}$ eV, and the hydrostatic pressure drops below 0.5 kbar. Relaxed structures and the calculated total energies are used when assessing the thermochemical stability  and as a starting point for defect calculations.

The metastability of the high-pressure pyrochlore phases was assessed in two ways. First, we used the equations of state that were computed for both chemistries in both structure types. Fixed volume relaxations of the cell-shape and ionic positions were performed for all systems on seven scaled structures with the scale of the linear dimension ranging from 0.97 to 1.03 in steps of 0.01. The numerical energy-volume dependence was then used to fit the Birch-Murnaghan equations of state \cite{Birch_PR:1947} (shown in Fig.\ \ref{fig:neb}a). The fitted equation of states (EOS) were then used to numerically determine the transition pressure from T- to P-phases, i.e., the slope of cotangent line, using convex hull analysis. The EOS were also used to calculate the enthalpy under external pressures as tabulated on Table S2 and S3.  Second, the transformation pathways for the concerted pyrochlore-thorvetite transformations were predicted using the solid-state nudged elastic band (SSNEB) method \cite{ssneb} with the initial pathways provided by our structure matching algorithm \cite{Stevanovic_PRM:2018,Therrien_JCP:2020}. A total of 19 intermediate images are used, each image containing 22 atoms to accommodate the unit cell of the pyrochlore structure. The above described numerical setup was used in SSNEB calculations. The pathway is considered converged when the forces on each image are less than 0.1 eV/\AA. Those images that appeared to represent local minima (intermediate states) along the pathway were relaxed separately to confirm that local minima had been found.

\subsection{Electronic structure}
To have a more accurate prediction of the electronic structure beyond the PBE exchange-correlation functional used for FERE method, we further applied perturbative many-body GW approximation to correct the eigenvalues. We followed the approach by Lany\cite{Peng_PRB_2013,Lany_PRB_2013}, in which PBE wavefunction are fixed while the eigenvalues are iteratively updated till the band gap difference is less than 0.1 eV.  
We set the number of unoccupied bands to be 95$\times$ the number atoms in the simulation cell to obtain converged GW eigenvalues. The Brillouin zone was sampled with $\Gamma$-center $k$-point grid with density of $N\times N_{kpts}$ = 2000, where $N$ is the number of atoms in the cell and $N_{kpts}$ is the number of $k$ points. We also ensured that the $k$-points associated with band edges are already included in the grid. 

We estimated the dispersion of band edge at PBE level by calculating the average band effective mass ($m^{*}_{b}$) for electron and hole. It was determined from the electronic density of states of a dense $k$-point grid assuming isotropic and parabolic band pockets\cite{Yan_EES_2015},
\begin{equation}
    m^{*}_{DOS} = N_{b}^{2/3}m^{*}_{b}
\end{equation}
where m$^{*}_{DOS}$ is the density-of-states effective mass and $N_{b}$ is the band degeneracy. The energy window of 100 meV from the relevant band edges was used.  The number of $k$-points ($n_{kpts}$) are determined by $N\times n_{kpts}$ = 8000, where $N$ is the number of atoms in the primitive cell. Gaussian smearing of 0.01 eV was used.

\subsection{Defects and doping}\label{sec:methods_defects}

To investigate the dopability of In$_{2}$X$_{2}$O$_{7}$ (X=Si, Ge) and identify their potential dopants, we calculated the defect formation energies ($\Delta E_{D,q}$) of their native defects (vacancy, interstitial, and antisite defects) and potential substitutional defects on both X and In site. Specifically, we utilized the well-established supercell approach\cite{Lany_MSMSE_2009, Goyal_CMS_2017},
\begin{equation}
\Delta E_{D,q} = E_{D,q} - E_{\text{bulk}} + \sum_{i} n_{i} \mu_{i} + qE_{\text{F}} + E_{\text{corr}}
\end{equation}
where $E_{D,q}$ and $E_{\text{bulk}}$ are the total energy of supercell with and without defects, respectively. $q$ is the charge state of the defect and $E_\mathrm{F}$ is the Fermi energy, which is referenced to the valence band maximum. 
$n_{i}$ is the number of atoms of element $i$ added to ($n_{i} < 0$) or removed from ($n_{i} > 0$) the bulk supercell to create the defect supercell. $\mu_{i}$ is the chemical potential of element $i$ and the range, which is evaluated via convex hull analysis, is reported in Table S4--7. $E_{\text{corr}}$ is the correction term that takes into account potential alignment between supercells with and without defect and the interaction between defects and their periodic images. The corrections were calculated using \textit{pylada-defects}.\cite{Goyal_PRMater_2018,Goyal_CMS_2017} The band gap is corrected by the GW method described in Section IIB. We applied a rigid shift since the wavefunction is fixed during the GW calculations and, consequently, the average electrostatic potential remains the same. 

For phase stability analysis, we note that T-In$_2$Si$_2$O$_7$, which is a experimentally stable structure under standard condition, is above the convex hull by 7 meV/atom using FERE. Considering the uncertainty of FERE at around 50 meV/atom and the experimental observation, we lowered the total energy of T-In$_2$Si$_2$O$_7$ by 30 meV/atom. We further examined the choice of the correction by comparing the predicted net carrier concentration as a function of temperature (see Figure S1) using different amount of correction (-7, -30, and -50 meV/atom). We found that the difference is within an order of magnitude and the conclusions of this paper remain qualitatively the same. 

To estimate the net donor concentration of Zr-doped In$_{2}$X$_{2}$O$_{7}$ (X=Si, Ge) in both structures, we utilized thermodynamic modeling \cite{Lany_JCP_2018}, with the carrier concentration estimated from the density-of-states effective mass and the defect concentrations (C$_{D}$) calculated using the corresponding defect formation energies ($\Delta E_{D}$) at equilibrium Fermi energy and chosen chemical potentials,
\begin{equation}
    C_{D}(,T)=C_{0}exp[\frac{-\Delta E_{D}}{kT}]
\end{equation}
where $C_{0}$ is the density of available lattice sites for the defect to form. 
Equilibrium Fermi energy is calculated self-consistently to obey charge neutrality among all the charged defects and charge carriers. 

\subsection{Thin film growth and characterization}
Thin films of UID and Zr-doped In$_2$Ge$_2$O$_7$ were grown on c-Al$_2$O$_3$ substrates by pulsed laser deposition using a KrF Excimer laser of wavelength 248 nm in a vacuum chamber with base pressure of 10$^{-8}$ Torr. The Zr-doped In$_2$Ge$_2$O$_7$ target with 5-10 percent excess GeO2 placed 7.5 cm from the substrate was ablated at a pulse repetition rate of 10 Hz and laser energies of 300 mJ. During deposition, the substrate heater setpoints were maintained at RT or 600°C and films were grown at a chamber oxygen pressure between 50-180 mTorr. The as-deposited films were annealed in laboratory ambient air in a box furnace for 1 hour at 800$^{\circ}$C.

The as-grown and post-deposition annealed thin film structure was analyzed by x-ray diffraction. A Bruker D8 Discover diffractometer equipped with Cu-K$\alpha$ radiation source and a 2D detector in a Bragg-Brentano configuration was used for all measurements. The surface morphology and roughness of deposited layers before and after postgrowth annealing was studied by atomic force microscopy (AFM). AFM images are acquired in tapping mode using a Veeco DI D3100 AFM equipped with a Nanoscope V controller.

The electrical properties of the annealed films are measured at room temperature using the van der Pauw configuration by Lakeshore fast hall equipment with a magnetic field of 1T. Before measurement, indium pads were pressed on the edges of the samples to improve contact resistance. The optical properties of the films were obtained by standard (isotropic) Spectroscopic Ellipsometry (SE) in the spectral range of 0.7–6.5 eV using a J.A Woollam Co. M-2000 ellipsometer. The ellipsometry amplitude ratio, $\Psi$ and phase difference, $\Delta$ spectra were measured at 55$^{\circ}$, 60$^{\circ}$, and 65$^{\circ}$ incident angles on homogenous as-grown and annealed In$_2$Ge$_2$O$_7$ films on c-Al$_2$O$_3$.

\section{Results and discussion}

\subsection{Synthesizability of ground-state T-phase}

\begin{figure}[!ht]
\centering
\includegraphics[width=0.9\linewidth]{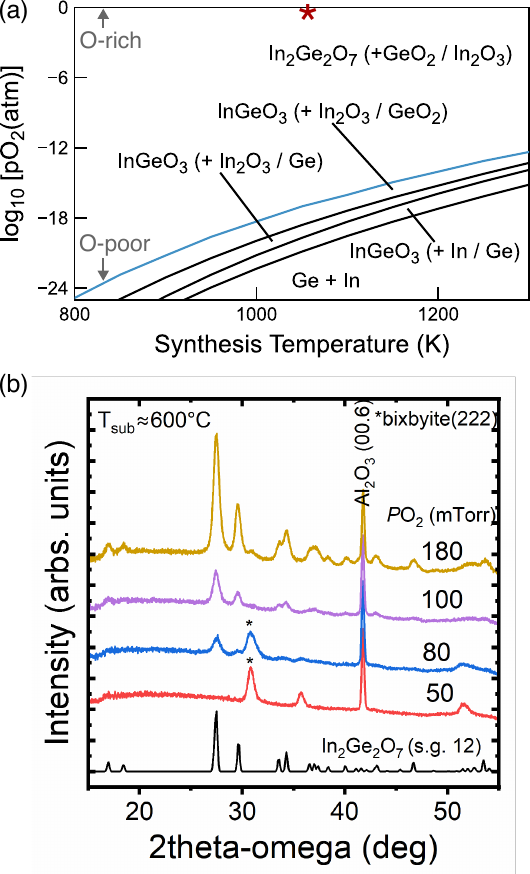}
\caption{\label{fig:pO2-T} 
Synthesizability of T-In$_2$Ge$_2$O$_7$. (a) Predicted partial oxygen pressure (pO$_2$) \textit{vs.} synthesis temperature phase diagrams for T-In$_2$Ge$_2$O$_7$. Parentheses indicate phases that can coexist when stoichiometry between In and Ge deviates from 1:1. Blue curve indicates the phase boundary for T-In$_2$Ge$_2$O$_7$ in oxygen-poor condition. Red asterisk indicates the nominal annealing condition in our experiments. (b) Post-Anneal XRD 2Theta-Omega scans of a series of T-In$_2$Ge$_2$O$_7$ thin films deposited on $\alpha$-Al$_2$O$_3$(00.1) substrates at 600$^{\circ}$C and pO$_2$ as indicated. All samples were annealed at 800$^{\circ}$C in air. The peaks marked with asterisks are associated with In$_2$O$_3$(222) and observed in samples grown at lower pO$_2$ .
}
\end{figure}

To help guide the synthesis of thin film for device application, we utilized the computational thermodynamic modeling approach we recently developed \cite{Lee_ACEELM_2024,Egbo_APL_2024,Callahan_APL_2024} to predict their pO$_2$-Temperature phase diagram (see Fig.\ \ref{fig:pO2-T} and Figure S2). The phase diagrams suggest that T-In$_2$Ge$_2$O$_7$ and T-In$_2$Si$_2$O$_7$ are the thermodynamic stable phase at the typical oxygen partial pressure (pO$_2 >$ 10$^{-10}$ atm) and synthesis temperature range ($<$ 1000$^{\circ}$C) for thin-film growth. Deviations from the 1:1 ratio between In and X (X = Si, Ge) will lead to coexistence with secondary phases --- In-rich synthesis condition can give rise to formation of In$_2$O$_3$ while X-rich condition can result in GeO$_2$ or SiO$_2$.   Careful tuning will be needed to avoid secondary phases. We noted that the expected uncertainty for the shown temperature range is less than an order of magnitude\cite{Lee_ACEELM_2024}. In the past, powders of T-In$_2$Si$_2$O$_7$ and T-In$_2$Ge$_2$O$_7$ were synthesized before by solid-state reactions\cite{gaewdang_structural_1994,li_comparative_2015,shannon_synthesis_1970,gaewdang_structural_1994, reid_high-pressure_1977}, but there are no reports of thin film growth of these materials in literature.

Fig.\ \ref{fig:pO2-T}(b) presents the X-ray diffraction (XRD) data for  annealed films deposited on c-Al$_2$O$_3$. Films grown at substrate temperature setpoints up to 600$^{\circ}$C exhibited amorphous structures (see Figure S3).  Post-growth annealing at 800$^{\circ}$C in laboratory ambient air led to the formation of a polycrystalline phase, with diffraction peaks corresponding to T-In$_2$Ge$_2$O$_7$ (space group 12). For the low oxygen partial pressures, bixbyite (222) peaks (denoted by *) were observed, particularly in films grown under lower O$_2$ partial pressure, indicating an In-rich composition for In$_2$Ge$_2$O$_7$. Notably, growth under O-deficient conditions at high temperatures resulted in the formation of as-grown In$_2$O$_3$ layers (reference peaks can be found in Fig.\ S3), suggesting that Ge-O species may be volatile under these conditions and desorb from the substrate. This is consistent with the predicted pO$_2$-Temperature phase diagram.  Overall, our results suggest that achieving single-phase In$_2$Ge$_2$O$_7$ growth requires careful control of Ge-O desorption. Further optimization of both growth and annealing conditions will be essential for the development of stoichiometric, single-phase, and high-quality crystalline T-In$_2$Ge$_2$O$_7$ thin films.

We further examined the surface morphology of the annealed thin films using atomic force microscopy (AFM) measurements. We evaluated the influence of composition by comparing representative stoichiometric In$_2$Ge$_2$O$_7$ thin films grown under O-rich conditions and In-rich In$_2$Ge$_2$O$_7$ thin films grown under O-poor conditions. As shown in Figure S4(b), the In-rich films exhibit a smoother surface characterized by small nanocrystals, a morphology commonly observed in In$_2$O$_3$-based thin films\cite{Bierwagen_JAP_2010}. In contrast, the stoichiometric films (Figure S4a) display a rougher surface with larger and  more randomly faceted grains.

\subsection{Electronic structure calculations and optical absorption measurements}

\begin{figure}
\includegraphics[width=0.90\linewidth]{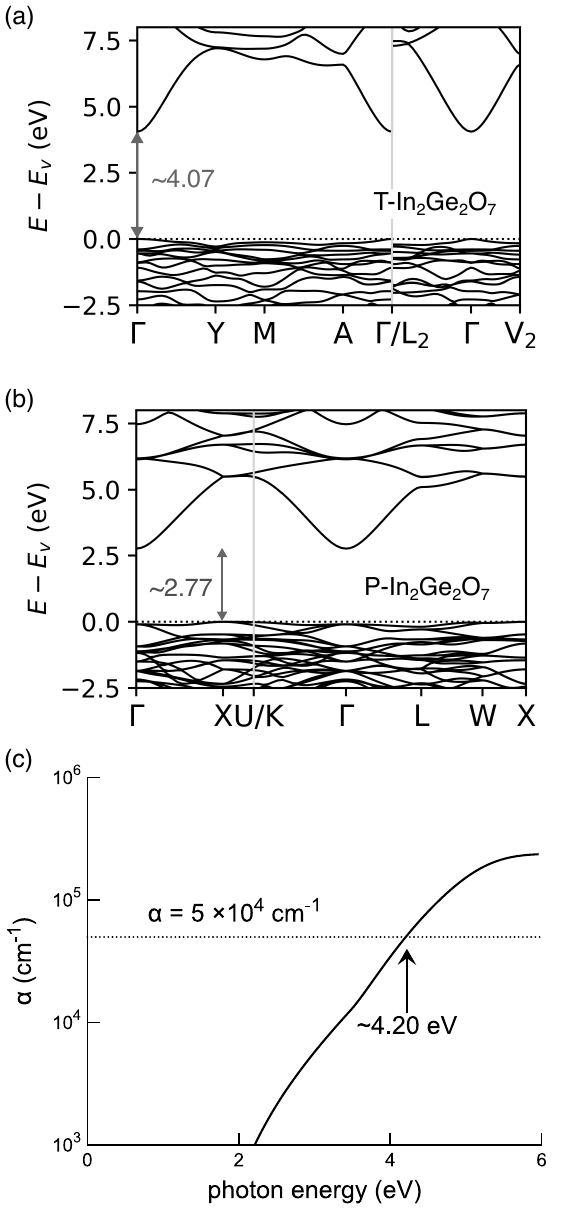}
\caption{\label{fig:es-Ge} Electronic structure for In$_2$Ge$_2$O$_7$. (a) and (b) are predicted electronic structure of T- and P- phases, respectively.  $E_v$ is the valence band maximum and GW correction was applied as a rigid scissor shift to valence and conduction band edges. T- and P-phases have direct and indirect band gap, respectively. (c) experimentally measured absorption spectra for UID T-In$_2$Ge$_2$O$_7$. 
}
\end{figure}

Fig.\ \ref{fig:es-Ge}(a) and (b) show the calculated electronic structure of T- and P-In$_2$Ge$_2$O$_7$, respectively (see Figure S5 for the ones for In$_2$Si$_2$O$_7$). They are based on PBE exchange-correlation functional corrected by the GW method. 
Table\ S8 tabulates the derived average effective mass and band degeneracy, along with dielectric and structural properties. For the same chemistry, we found that thorvetite structure has higher electronic band gap than its pyrochlore counterpart and as a consequence has lower electronic contribution to static dielectric constant. Additionally,  we found that ionic contribution accounts for majority part of the static dielectric constant and the pyrochlore structure is more polarizable. Together, the pyrochlore structure generally has larger static dielectric constant than the thorvetite counterpart.  Carrier mobility can be affected by extrinsic factors and the strength of electron-phonon coupling but generally depends on the carrier effective mass. We find that the pyrochlore structure generally has larger electron effective mass but smaller hole effective mass than the thorvetite one. 

\begin{figure*}[!bht]
\centering
\includegraphics[width=0.95\linewidth]{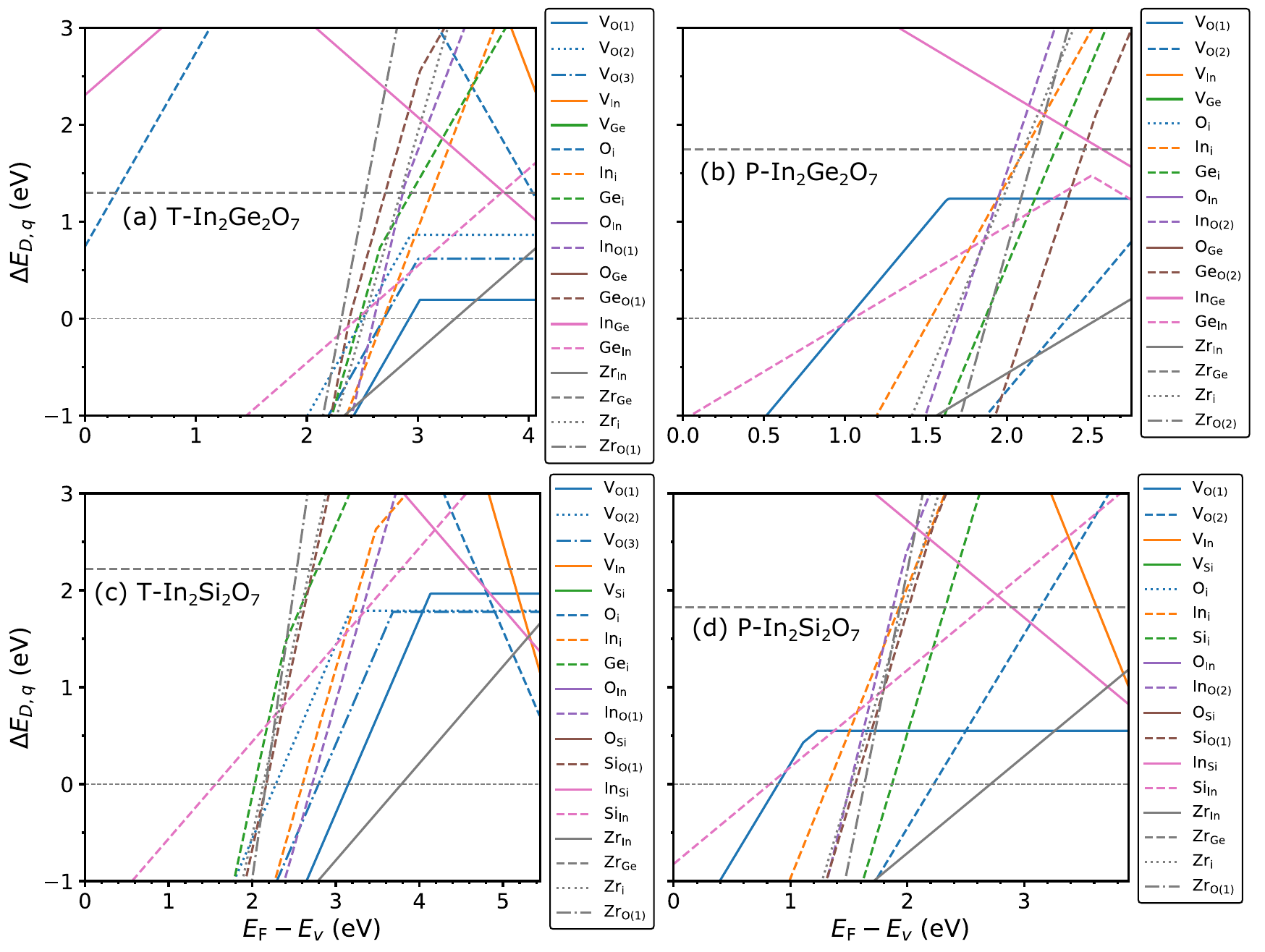}
\caption{\label{fig2} 
Defect formation energy plots of all 4 studied compounds are shown under O-poor condition (Vertices V1 on Table\ S5--8) Lines represent defect formation energies as a function of the Fermi energy (expressed relative to the valence band maximum, $E_v$ of every compound) for all native defects and Zr impurities. Defect plots for the high pressure P-In$_{2}$Ge$_{2}$O$_{2}$ and P-In$_{2}$Si$_{2}$O$_{2}$ phases are evaluated at 6 GPa and 12 GPa, respectively (see text for details). O(1), O(2), and O(3) indicate different Wyckoff sites of oxygen atoms. 
}
\end{figure*}

These electronic structure calculation results are important because the calculated properties play a key role in the performance of a material in power electronics. Baliga figure of merit (BFOM) quantifies the inverse of the theoretical on-state conduction loss and is defined as \cite{baliga_semiconductors_1982},
\begin{equation}
    \mathrm{BFOM} = \epsilon \mu E_b^{3}
\end{equation}
where $\epsilon$, $\mu$, and $E_b$ are static dielectric constant, carrier mobility, and dielectric breakdown field, respectively. The latter two and the electronic contribution to dielectric constant are all closely related to electronic structure.  Overall, the thorvetite structure is expected to have better n-type BFOM since BFOM scales with $E_b$ to the power of 3, which is empirically equivalent to power of 6 scaling with band gap.\cite{tsao_ultrawide-bandgap_2018} Comparing Ge to Si, we found that Si has larger band gap but smaller static dielectric constant and hole effective mass. Si and Ge have comparable electron effective mass. Since band gap dominates the BFOM, In$_2$Si$_2$O$_7$ has been predicted to be a better power electronics for low-frequency application than In$_2$Ge$_2$O$_7$ when the same structure is considered.

Fig.\ \ref{fig:es-Ge}(c) shows the absorption spectra for unintentionally doped (UID) T-In$_2$Ge$_2$O$_7$ thin film, deduced from spectroscopic ellipsometry measurements, of n and k optical constants. Using absorption onset of 5$\times$10$^4$ cm$^{-1}$, optical band gap was found to be around 4.2 eV, and assuming the absorption onset of  1$\times$10$^4$ cm$^{-1}$ the estimated optical band gap is just under  4 eV. The measured band gap is comparable with the predicted bandgap of $\sim$4.1 eV.  There is an additional sub-gap absorption in the experimental spectra, which is likely attributed to defect absorption due to polycrystalline morphology of the synthesized and air-annealed samples.

\subsection{Defect and dopant calculations}

To investigate behavior of intrinsic defects and extrinsic dopants we performed point defect calculations for all 4 compounds studied here. Defect formation energy plots are shown in Fig.~\ref{fig2} for both T- and P- phases of In$_{2}$Ge$_{2}$O$_{2}$ and In$_{2}$Si$_{2}$O$_{2}$. As noted previously, our focus is on $n$-type dopability of these compounds. The defect plots are hence, shown for the most oxygen poor conditions within their respective stability domains in the chemical potential spaces (lowest $\Delta \mu_{\mbox{O}}$, see Table S4--S7) as those are the most favorable for achieving the highest concentrations of donor defects. Defect plots corresponding to the O-rich conditions are shown in Figure S6. Also, for the high-pressure P- phases the defect plots are evaluated at pressures at which the respective phase is lower in enthalpy than their low pressure T- counterparts as explained in Section~\ref{sec:methods_stability}. Pressures of 6 and 12 GPa were chosen for P-In$_{2}$Ge$_{2}$O$_{2}$ and P-In$_{2}$Si$_{2}$O$_{2}$, respectively, based on the reported experimental synthesis conditions. We also ensure that the values are larger than their respective transition pressure.

All intrinsic point defects, including vacancies of all constituent atoms, their interstitial and antisite defects are considered and their formation energies are shown in Fig.~\ref{fig2}. In addition, the formation energies of Zr as an extrinsic impurity (dopant) are evaluated for different substitutional and interstitial sites. As evident from Fig.~\ref{fig2}, all relevant, i.e., low-energy, intrinsic defects at O-poor conditions are donor-behaving defects. These include O-vacancies, In and Ge interstitials, as well as Ge$_{\text{In}}$, Ge$_\text{O}$ and In$_\text{O}$ antisite defects. This is true for both In$_{2}$Ge$_{2}$O$_{2}$  and In$_{2}$Si$_{2}$O$_{2}$ in both the crystal structures. In addition, substitutional Zr$_\text{In}$, Zr$_\text{O}$ as well as Zr$_\text{i}$ defects also behave as donors, while, as expected,  Zr$_\text{Ge}$ is electrically neutral (does not create charge carriers of any kind). In all of these systems the lowest energy donor defect for Fermi levels ($E_{\mathrm{F}}$) close to the conduction band is the Zr$_\text{In}$ substitutional impurity in +1 charge state (donates one electron per defect) and there are virtually no acceptor defects with low enough energy to obstruct $n$-type behavior. All this renders In$_{2}$Ge$_{2}$O$_{2}$  and In$_{2}$Si$_{2}$O$_{2}$ intrinsically $n$-type dopable, in both T- and P- structures, with Zr as a possible candidate extrinsic dopant. These predictions pertain to O-poor conditions, which imply oxygen partial pressures of $<$ 10$^{-12}$ atm at temperatures between 800 and 1300 K as shown in Fig.~\ref{fig:pO2-T}. In addition to Zr, we also explored other possible substitutional donor dopants in +1 charge state using P-In$_2$Ge$_2$O$_7$ as an representative host material. We considered elements with nominal 4+ charge state substituting for In site and elements with nominal 5+ charge state substituting for Ge site. Specifically, only the ones with comparable ionic radius to respective substituents are considered. Their defect formation energies when $E_{\mathrm{F}}$ is at valence band maximum are tabulated on Table S9. Zr substitutional defect remains the most promising one with lowest formation energy, closely followed by Hf substitutional defects. As$_{\mathrm{Ge}}^{+1}$ defect can be another potential choice but its formation energy is higher by $\sim$0.2 eV than Zr$_\mathrm{In}^{+1}$.

Various acceptor defects, such as V$_\text{In}$,  V$_\text{Ge}$, and O$_\text{i}$ become more relevant at O-rich conditions (see Figure S6). More specifically, V$_\text{In}$ and In$_\text{Si}$ or In$_\text{Ge}$ appear as lowest energy acceptors for all 4 systems at O-rich conditions and for E$_\text{F}$ closer to the conduction band minimum (CBM). Interstitial O is a low energy acceptor at the same conditions only for the ground state T- structures, for which  O antisites as well as V$_\text{Si}$ or V$_\text{Ge}$ also appear low energy acceptors for $\text{E}_\text{F} \lesssim \text{CBM}$. For most of these acceptors, their formation energies becomes negative (exothermic) above a certain E$_\text{F}$ value that is still below the CBM. Hence, contrary to O-poor conditions, these defects would largely compensate the attempt to introduce free electrons to all 4 compounds, rendering them lightly (In$_2$Ge$_2$O$_7$) or not (In$_2$Si$_2$O$_7$) $n$-type dopable at O-rich conditions. 

\subsection{Calculated and measured doping}
\begin{figure}[!hbt]
\centering
\includegraphics[width=0.95\linewidth]{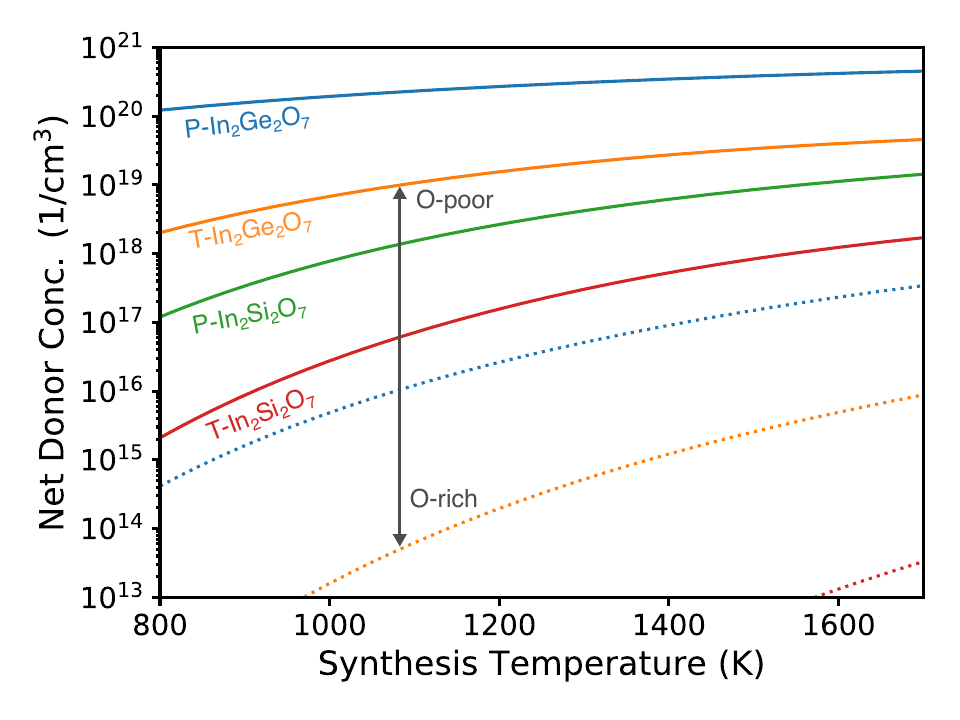}
\caption{\label{fig:conc} 
Predicted Net donor concentration for Zr-doped In$_2$X$_2$O$_7$ (X=Si,Ge) as a function of synthesis temperature under O-poor (solid) and O-rich (dotted) conditions. The double-headed arrow indicates the range for T-In$_2$Ge$_2$O$_7$ at synthesis temperature $\sim$1073 K.
}
\end{figure}

Fig.\ \ref{fig:conc} shows the predicted net donor concentration for Zr-doped In$_2$X$_2$O$_7$ (X=Si,Ge) at different synthesis temperatures. Reflecting our discussion of defect formation energy, the net donor concentration under O-poor condition can be larger than the concentration under O-rich condition by approximately 5 orders of magnitude at the studied temperature range. For Zr-doped P-In$_2$Si$_2$O$_7$ at O-rich condition, the net donor concentration is comparable to intrinsic carrier concentrations and can even become p-type at synthesis temperature $\gtrsim$ 1000 K. 

For the ideal synthesis condition for n-type doping (O-poor), we find that P-phase has higher net donor concentration than T-phase within the same chemistry. We also find that In$_2$Ge$_2$O$_7$ in both phases has higher net donor concentration than  In$_2$Si$_2$O$_7$. Overall, the trend for net carrier concentration is associated with the band gap --- T-In$_2$Si$_2$O$_7$ has the largest gap but lowest net carrier concentration; in contrast, P-In$_2$Ge$_2$O$_7$ has the smallest gap but highest net carrier concentration. We also notice that O interstitials only act as the dominant charge compensating acceptor defects to Zr-doping in T-phases. The energetics of O interstitials are higher in P-phases and this is consistent with our intuition that high-pressure phases have smaller volume than low-pressure phases and thus, it is harder for O interstitials to form.

Table\ \ref{table:expt} summarizes the electrical properties of undoped (Z0) and Zr-doped (Z1–Z3) T-In$_2$Ge$_2$O$_7$ thin films following post-growth annealing at 800$^{\circ}$C. The atomic concentration of Zr dopants and the Ge/In composition, as determined by X-ray fluorescence, are also presented. All undoped T-In$_2$Ge$_2$O$_7$ films exhibited high resistivity ($>$3 M$\Omega \cdot$cm), indicating their insulating nature. In contrast, Zr doping introduced n-type conductivity, with carrier concentrations ranging from 10$^{14}$ to 10$^{16}$ cm$^{-3}$. Given that the samples are annealed in ambient air, the measured carrier concentrations are consistent with predicted values under the most O-rich condition ($\approx$10$^{14}$ cm$^{-3}$ at 800$^{\circ}$C). Growing samples at lower oxygen partial pressure is the key to improving carrier concentration in the future. In addition, all doped films displayed low carrier mobility ($\sim$1 $\frac{\mathrm{cm}^{2}}{V\cdot s}$). To enhance carrier transport in Zr-doped T-In$_2$Ge$_2$O$_7$ thin films, further optimization of key factors such as microstructure and strain is necessary. These improvements will be critical for achieving high-mobility n-type conductivity in this novel ternary oxide material.
\begin{table}[!ht]
\caption{Comparison of room temperature electrical properties of undoped and Zr-doped T-In$_2$Ge$_2$O$_7$ thin films. $\mu$, $n$, and $\rho$ are mobility, electron concentration, and resistivity, respectively. } 
    \begin{tabular}{ccccccc}
    \hline
    & \\
    Sample & Zr(\%) & Ge(\%) & In(\%) & $\mu$ ($\frac{\mathrm{cm}^2}{V\cdot s}$) & $n$ (1/cm$^{3}$) & $\rho$ (k$\Omega \cdot$cm)  \\ 
    & \\
    \hline 
    & \\
    Z0 & - & 52.1 & 47.9 & - & - & $>$3000  \vspace{0.2 cm} \\
    Z1 & 1.25 & 51.3 & 47.4 & 0.2 & 2.40$\times$10$^{15}$ & 13  \vspace{0.2 cm} \\ 
    Z2 & 1.36 & 48.0 & 50.3 & 0.59 & 6.50$\times$10$^{14}$ & 50  \vspace{0.2 cm} \\ 
    Z3 & 1.76 & 52.4 & 46.0 & 1.1 & 3.70$\times$10$^{16}$ & 0.15 \vspace{0.2 cm}  \\ \hline
    \end{tabular}
\label{table:expt}
\end{table}

\subsection{Thermodynamics and kinetics of T- and P- phase transitions}
\begin{figure}[!htb]
\includegraphics[width=\linewidth]{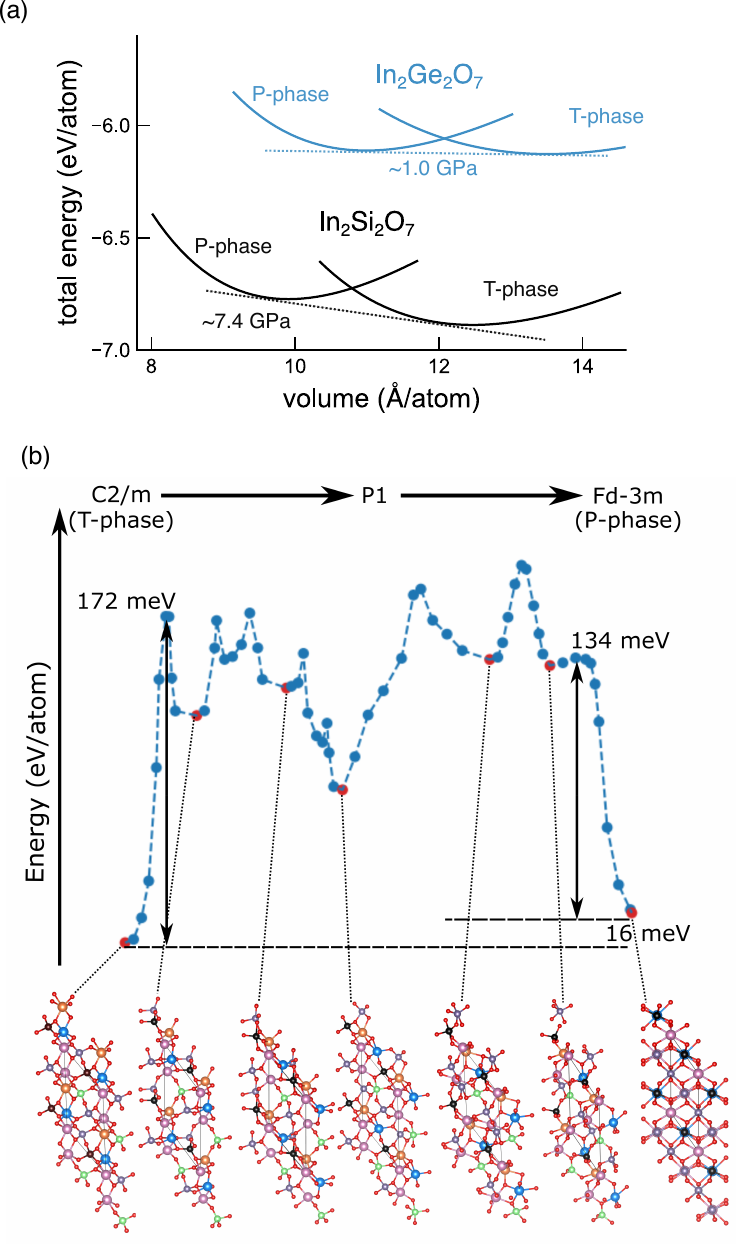}
\caption{\label{fig:neb} Phase transition for In$_2$X$_2$O$_7$ (X=Si,Ge). (a) Predicted equations of states. The transition pressures derived from cotangent lines are shown.   (b) Predicted minimum energy path for the transformation from thortveitite to pyrochlore in In$_2$Ge$_2$O$_7$ (see text for details). Purple, green, and black atoms are Ge; Pink, blue, and orange atoms are In; Red atoms are O. 
}
\end{figure}

High-pressure pyrochlore phases of In$_2$X$_2$O$_7$ (X=Si,Ge) are also promising wide-gap semiconductors and understanding their thermodynamics is key for future experimental realization. 
Fig.\ \ref{fig:neb}(a) shows the predicted equation of states (EOS) for their T- and P-phases. The results support the existing knowledge that P-phases are the high-pressure phases since P-phases have higher total energy per atom but lower atomic volume than T-phases. The transition pressure between the two phases are determined by the equal enthalpy condition and graphically by the negative slope of the cotangent lines between the EOS of the two phases. We predicted that the transition pressures from T- to P-phases are $\sim$1.0 GPa and $\sim$7.4 GPa for In$_2$Ge$_2$O$_7$ and In$_2$Si$_2$O$_7$, respectively. Positive transition pressure indicates external pressure is required to stabilize P-phases instead of T-phases.
Fig.\ \ref{fig:neb}(a) also shows that P-In$_2$Si$_2$O$_7$ requires higher external pressure than P-In$_2$Ge$_2$O$_7$, which is consistent with reported synthesis conditions.

High-pressure pyrochlore phases are experimentally observed to be long-lived metastable states but the kinetics of the phase transformation remains unclear.
The transformation path is highly non-trivial since the atomic structures of these two phases have significant differences in local coordination environments and the relative arrangement of coordination polyhedra. The thortveitite structure has a primitive cell containing one formula unit, with 4 equivalent 6-fold coordinated In atoms and 2 equivalent tetrahedrally coordinated Ge or Si atoms. In this structure, there are alternating planes of In and Ge or Si cations with miller index (100). The pyrochlore structure has a primitive cell consisting of 2 formula units, with 8-fold coordinated In atoms and 6-fold coordinated Ge atoms. This structure has no family of miller planes for which all planes contain only one cation, instead there are tetrahedra consisting of 4 In atoms (Wyckoff sites c) neighboring tetrahedra consisting of 4 Ge or Si atoms (Wyckoff sites d). Herein, we propose a possible transformation pathway between these phases using the solid-state nudged elastic band method, starting from an optimal atom-to-atom map between the initial and final structures. 

Fig.\ \ref{fig:neb}(b) shows the (bulk) transformation pathway between the thortveitite and pyrochlore structures for In$_2$Ge$_2$O$_7$. There are five distinct local minima between the initial and final structures. In the T-phase to P-phase direction, the rate limiting step is between the T-phase and the first intermediate local minimum, with a barrier of 172 meV/atom, while in the P-phase to T-phase direction, the rate limiting step is that from the P-phase to the first intermediate local minimum with a barrier of 134 meV/atom. All intermediate images have space group P(1). The energetic landscape of phase transformation for In$_2$Si$_2$O$_7$ is similar, with higher barrier for the rate-limiting step from T-Phase to P-phase. The barrier in the reverse direction is lower for In$_2$Si$_2$O$_7$ due to relatively higher energy of the P-phase (see Fig.\ \ref{fig:neb}a). Using In$_2$Ge$_2$O$_7$ as example, the energetic penalties along the pathway are primarily associated with changes in site for In and Ge atoms. The first step requires diffusion and reduced coordination of two In atoms per cell (orange and blue). The second step involves change in oxygen coordination of an In and Ge atom (blue and black) in each cell. The blue In atom also inverts its polyhedron across a plane of oxygen atoms during the second step. The third step involves site changes of two In atoms and a Ge atom (blue, orange, and purple) in each cell. The fourth, fifth and sixth steps involve significant changes in volume and increases in coordination of all In and Ge atoms. 

The height of the reverse barrier suggests that, once formed, the pyrochlore structure is unlikely to transform back to the thortveitite structure at ambient conditions. The volume of the image with the greatest energy is greater than the volume of either the thortveitite or pyrochlore structures. Therefore, increasing pressure will increase the enthalpy barrier rather than lowering it, and for this pathway, high temperatures are necessary to enable the transformation between these structures.  These estimates are consistent with the experimentally reported synthesis conditions of 
moderately high pressures of 5.2-6.5 GPa  and synthesis temperatures of 1373-1573 K for In$_2$Ge$_2$O$_7$.\cite{li_comparative_2015, shannon_synthesis_1968, messous_indium_1995}


\section{Conclusion}
In summary, we conducted a systematic investigation into one of the most promising predicted chemistry groups, In$_2$X$_2$O$_7$ (X=Si,Ge), for power electronics.
Our previous search was based on \textit{intrinsic} properties like the Baliga Figure of merit and thermal conductivity while in this paper we focused on evaluating their potential as active materials for a functional device. To this end, we synergically performed computational and experimental studies on their synthesizability, electronic structure, and doping tendency. We not only predicted that all four In$_2$X$_2$O$_7$ compounds are n-type dopable under O-poor conditions but also successfully grew Zr-doped T-In$_2$Ge$_2$O$_7$ thin films on $\alpha$-Al$_2$O$_3$ (00.1) substrate to support the prediction. This study validates one of the top candidates from previous works and establishes In$_2$X$_2$O$_7$ as promising n-type wide-gap semiconductors. Driven by the high theoretical net donor concentration at O-poor condition, our future work will focus on optimizing the synthesis process to make In$_2$X$_2$O$_7$-based devices with higher carrier concentration and mobility.

\begin{acknowledgments}
This work was authored in part at the National Renewable Energy Laboratory (NREL) operated by Alliance for Sustainable Energy, LLC, for the U.S. Department of Energy (DOE) under Contract No. DE-AC36-08GO28308. Funding provided by the Laboratory Directed Research and Development (LDRD) program at NREL. This work was primarily supported as part of APEX (A Center for Power Electronics Materials and Manufacturing Exploration), an Energy Frontier Research Center funded by the U.S. Department of Energy, Office of Science, Basic Energy Sciences under Award \# ERW0345. H.G.'s internship at NREL (electrical doping measurements) was made possible by the NREL STAR internship program. We would like to thank Prof. Stephanie Karen Hurst for useful discussions with H.G. and the rest of the experimental team. Experimental band gap measurements were supported by NREL LDRD program. Computational research was performed using computational resources sponsored by the Department of Energy’s Office of Energy Efficiency and Renewable Energy, located at the National Renewable Energy Laboratory.
The authors also acknowledge Colorado School of Mines supercomputing resources (http://ciarc.mines.edu/hpc) made available for conducting the research reported in this paper.
The views expressed in the article do not necessarily represent the views of the DOE or the U.S. Government.\\
\end{acknowledgments}

\noindent
\textbf{Conflict of Interest}\\
The authors have no conflicts to disclose.\\

\noindent
\textbf{Dara Availability}\\
Data associated with this study are available from the corresponding authors upon request.

\bibliography{In2Ge2O7}

\begin{thebibliography}{40}%
\makeatletter
\providecommand \@ifxundefined [1]{%
 \@ifx{#1\undefined}
}%
\providecommand \@ifnum [1]{%
 \ifnum #1\expandafter \@firstoftwo
 \else \expandafter \@secondoftwo
 \fi
}%
\providecommand \@ifx [1]{%
 \ifx #1\expandafter \@firstoftwo
 \else \expandafter \@secondoftwo
 \fi
}%
\providecommand \natexlab [1]{#1}%
\providecommand \enquote  [1]{``#1''}%
\providecommand \bibnamefont  [1]{#1}%
\providecommand \bibfnamefont [1]{#1}%
\providecommand \citenamefont [1]{#1}%
\providecommand \href@noop [0]{\@secondoftwo}%
\providecommand \href [0]{\begingroup \@sanitize@url \@href}%
\providecommand \@href[1]{\@@startlink{#1}\@@href}%
\providecommand \@@href[1]{\endgroup#1\@@endlink}%
\providecommand \@sanitize@url [0]{\catcode `\\12\catcode `\$12\catcode
  `\&12\catcode `\#12\catcode `\^12\catcode `\_12\catcode `\%12\relax}%
\providecommand \@@startlink[1]{}%
\providecommand \@@endlink[0]{}%
\providecommand \url  [0]{\begingroup\@sanitize@url \@url }%
\providecommand \@url [1]{\endgroup\@href {#1}{\urlprefix }}%
\providecommand \urlprefix  [0]{URL }%
\providecommand \Eprint [0]{\href }%
\providecommand \doibase [0]{https://doi.org/}%
\providecommand \selectlanguage [0]{\@gobble}%
\providecommand \bibinfo  [0]{\@secondoftwo}%
\providecommand \bibfield  [0]{\@secondoftwo}%
\providecommand \translation [1]{[#1]}%
\providecommand \BibitemOpen [0]{}%
\providecommand \bibitemStop [0]{}%
\providecommand \bibitemNoStop [0]{.\EOS\space}%
\providecommand \EOS [0]{\spacefactor3000\relax}%
\providecommand \BibitemShut  [1]{\csname bibitem#1\endcsname}%
\let\auto@bib@innerbib\@empty
\bibitem [{\citenamefont {Blaabjerg}\ \emph {et~al.}(2023)\citenamefont
  {Blaabjerg}, \citenamefont {Yang}, \citenamefont {Kim},\ and\ \citenamefont
  {Rodriguez}}]{blaabjerg_power_2023}%
  \BibitemOpen
  \bibfield  {author} {\bibinfo {author} {\bibfnamefont {F.}~\bibnamefont
  {Blaabjerg}}, \bibinfo {author} {\bibfnamefont {Y.}~\bibnamefont {Yang}},
  \bibinfo {author} {\bibfnamefont {K.~A.}\ \bibnamefont {Kim}},\ and\ \bibinfo
  {author} {\bibfnamefont {J.}~\bibnamefont {Rodriguez}},\ }\bibfield  {title}
  {\bibinfo {title} {Power {{Electronics Technology}} for {{Large-Scale
  Renewable Energy Generation}}},\ }\href
  {https://doi.org/10.1109/JPROC.2023.3253165} {\bibfield  {journal} {\bibinfo
  {journal} {Proceedings of the IEEE}\ }\textbf {\bibinfo {volume} {111}},\
  \bibinfo {pages} {335} (\bibinfo {year} {2023})}\BibitemShut {NoStop}%
\bibitem [{IEC(2023)}]{IEC_power_2023}%
  \BibitemOpen
  \href@noop {} {\emph {\bibinfo {title} {Power Semiconductors for an
  Energy-Wise Society}}},\ \bibinfo {type} {Tech. Rep.}\ (\bibinfo
  {institution} {International Electrotechnical Commission},\ \bibinfo
  {address} {Geneva, Switzerland},\ \bibinfo {year} {2023})\BibitemShut
  {NoStop}%
\bibitem [{\citenamefont {Kaplar}\ \emph {et~al.}(2017)\citenamefont {Kaplar},
  \citenamefont {Neely}, \citenamefont {Huber},\ and\ \citenamefont
  {Rashkin}}]{kaplar_generation-after-next_2017}%
  \BibitemOpen
  \bibfield  {author} {\bibinfo {author} {\bibfnamefont {R.~J.}\ \bibnamefont
  {Kaplar}}, \bibinfo {author} {\bibfnamefont {J.~C.}\ \bibnamefont {Neely}},
  \bibinfo {author} {\bibfnamefont {D.~L.}\ \bibnamefont {Huber}},\ and\
  \bibinfo {author} {\bibfnamefont {L.~J.}\ \bibnamefont {Rashkin}},\
  }\bibfield  {title} {\bibinfo {title} {Generation-{{After-Next Power
  Electronics}}: {{Ultrawide-bandgap}} devices, high-temperature packaging, and
  magnetic nanocomposite materials},\ }\href
  {https://doi.org/10.1109/MPEL.2016.2643098} {\bibfield  {journal} {\bibinfo
  {journal} {IEEE Power Electronics Magazine}\ }\textbf {\bibinfo {volume}
  {4}},\ \bibinfo {pages} {36} (\bibinfo {year} {2017})}\BibitemShut {NoStop}%
\bibitem [{\citenamefont {Garrity}\ \emph {et~al.}(2022)\citenamefont
  {Garrity}, \citenamefont {Lee}, \citenamefont {Gorai}, \citenamefont
  {Tellekamp}, \citenamefont {Zakutayev},\ and\ \citenamefont
  {Stevanovi\ifmmode~\acute{c}\else \'{c}\fi{}}}]{Garrity_PRXEnergy_2022}%
  \BibitemOpen
  \bibfield  {author} {\bibinfo {author} {\bibfnamefont {E.~M.}\ \bibnamefont
  {Garrity}}, \bibinfo {author} {\bibfnamefont {C.-W.}\ \bibnamefont {Lee}},
  \bibinfo {author} {\bibfnamefont {P.}~\bibnamefont {Gorai}}, \bibinfo
  {author} {\bibfnamefont {M.~B.}\ \bibnamefont {Tellekamp}}, \bibinfo {author}
  {\bibfnamefont {A.}~\bibnamefont {Zakutayev}},\ and\ \bibinfo {author}
  {\bibfnamefont {V.}~\bibnamefont {Stevanovi\ifmmode~\acute{c}\else
  \'{c}\fi{}}},\ }\bibfield  {title} {\bibinfo {title} {Computational
  identification of ternary wide-band-gap oxides for high-power electronics},\
  }\href {https://doi.org/10.1103/PRXEnergy.1.033006} {\bibfield  {journal}
  {\bibinfo  {journal} {PRX Energy}\ }\textbf {\bibinfo {volume} {1}},\
  \bibinfo {pages} {033006} (\bibinfo {year} {2022})}\BibitemShut {NoStop}%
\bibitem [{\citenamefont {Gorai}\ \emph {et~al.}(2019)\citenamefont {Gorai},
  \citenamefont {McKinney}, \citenamefont {Haegel}, \citenamefont {Zakutayev},\
  and\ \citenamefont {Stevanovic}}]{gorai_computational_2019}%
  \BibitemOpen
  \bibfield  {author} {\bibinfo {author} {\bibfnamefont {P.}~\bibnamefont
  {Gorai}}, \bibinfo {author} {\bibfnamefont {R.~W.}\ \bibnamefont {McKinney}},
  \bibinfo {author} {\bibfnamefont {N.~M.}\ \bibnamefont {Haegel}}, \bibinfo
  {author} {\bibfnamefont {A.}~\bibnamefont {Zakutayev}},\ and\ \bibinfo
  {author} {\bibfnamefont {V.}~\bibnamefont {Stevanovic}},\ }\bibfield  {title}
  {\bibinfo {title} {A computational survey of semiconductors for power
  electronics},\ }\href {https://doi.org/10.1039/C9EE01529A} {\bibfield
  {journal} {\bibinfo  {journal} {Energy \& Environmental Science}\ }\textbf
  {\bibinfo {volume} {12}},\ \bibinfo {pages} {3338} (\bibinfo {year}
  {2019})}\BibitemShut {NoStop}%
\bibitem [{\citenamefont {Reese}\ \emph {et~al.}(2019)\citenamefont {Reese},
  \citenamefont {Remo}, \citenamefont {Green},\ and\ \citenamefont
  {Zakutayev}}]{reese_how_2019}%
  \BibitemOpen
  \bibfield  {author} {\bibinfo {author} {\bibfnamefont {S.}~\bibnamefont
  {Reese}}, \bibinfo {author} {\bibfnamefont {T.}~\bibnamefont {Remo}},
  \bibinfo {author} {\bibfnamefont {J.}~\bibnamefont {Green}},\ and\ \bibinfo
  {author} {\bibfnamefont {A.}~\bibnamefont {Zakutayev}},\ }\bibfield  {title}
  {\bibinfo {title} {How {Much} {Will} {Gallium} {Oxide} {Power} {Electronics}
  {Cost}?},\ }\href@noop {} {\bibfield  {journal} {\bibinfo  {journal} {Joule}\
  }\textbf {\bibinfo {volume} {3}},\ \bibinfo {pages} {899} (\bibinfo {year}
  {2019})}\BibitemShut {NoStop}%
\bibitem [{\citenamefont {Garrity}\ \emph {et~al.}(2023)\citenamefont
  {Garrity}, \citenamefont {Egbo}, \citenamefont {Lee}, \citenamefont
  {Zakutayev},\ and\ \citenamefont {Stevanovi{\'c}}}]{Garrily_CM_2023_IIIBO3}%
  \BibitemOpen
  \bibfield  {author} {\bibinfo {author} {\bibfnamefont {E.~M.}\ \bibnamefont
  {Garrity}}, \bibinfo {author} {\bibfnamefont {K.}~\bibnamefont {Egbo}},
  \bibinfo {author} {\bibfnamefont {C.-W.}\ \bibnamefont {Lee}}, \bibinfo
  {author} {\bibfnamefont {A.}~\bibnamefont {Zakutayev}},\ and\ \bibinfo
  {author} {\bibfnamefont {V.}~\bibnamefont {Stevanovi{\'c}}},\ }\bibfield
  {title} {\bibinfo {title} {Defect chemistry and doping of ultrawide band gap
  (iii)bo3 compounds},\ }\href {https://doi.org/10.1021/acs.chemmater.3c01754}
  {\bibfield  {journal} {\bibinfo  {journal} {Chem. Mater.}\ }\textbf {\bibinfo
  {volume} {35}},\ \bibinfo {pages} {9952} (\bibinfo {year}
  {2023})}\BibitemShut {NoStop}%
\bibitem [{\citenamefont {Baliga}(1982)}]{baliga_semiconductors_1982}%
  \BibitemOpen
  \bibfield  {author} {\bibinfo {author} {\bibfnamefont {B.~J.}\ \bibnamefont
  {Baliga}},\ }\bibfield  {title} {\bibinfo {title} {Semiconductors for
  high‐voltage, vertical channel field‐effect transistors},\ }\href@noop {}
  {\bibfield  {journal} {\bibinfo  {journal} {Journal of Applied Physics}\
  }\textbf {\bibinfo {volume} {53}},\ \bibinfo {pages} {1759} (\bibinfo {year}
  {1982})}\BibitemShut {NoStop}%
\bibitem [{\citenamefont {Chae}\ \emph {et~al.}(2019)\citenamefont {Chae},
  \citenamefont {Lee}, \citenamefont {Mengle}, \citenamefont {Heron},\ and\
  \citenamefont {Kioupakis}}]{Chae_APL_2019}%
  \BibitemOpen
  \bibfield  {author} {\bibinfo {author} {\bibfnamefont {S.}~\bibnamefont
  {Chae}}, \bibinfo {author} {\bibfnamefont {J.}~\bibnamefont {Lee}}, \bibinfo
  {author} {\bibfnamefont {K.~A.}\ \bibnamefont {Mengle}}, \bibinfo {author}
  {\bibfnamefont {J.~T.}\ \bibnamefont {Heron}},\ and\ \bibinfo {author}
  {\bibfnamefont {E.}~\bibnamefont {Kioupakis}},\ }\bibfield  {title} {\bibinfo
  {title} {Rutile geo2: An ultrawide-band-gap semiconductor with ambipolar
  doping},\ }\href {https://doi.org/10.1063/1.5088370} {\bibfield  {journal}
  {\bibinfo  {journal} {Applied Physics Letters}\ }\textbf {\bibinfo {volume}
  {114}},\ \bibinfo {pages} {102104} (\bibinfo {year} {2019})}\BibitemShut
  {NoStop}%
\bibitem [{\citenamefont {Mengle}\ \emph {et~al.}(2019)\citenamefont {Mengle},
  \citenamefont {Chae},\ and\ \citenamefont {Kioupakis}}]{Mengle_JAP_2019}%
  \BibitemOpen
  \bibfield  {author} {\bibinfo {author} {\bibfnamefont {K.~A.}\ \bibnamefont
  {Mengle}}, \bibinfo {author} {\bibfnamefont {S.}~\bibnamefont {Chae}},\ and\
  \bibinfo {author} {\bibfnamefont {E.}~\bibnamefont {Kioupakis}},\ }\bibfield
  {title} {\bibinfo {title} {Quasiparticle band structure and optical
  properties of rutile geo2, an ultra-wide-band-gap semiconductor},\ }\href
  {https://doi.org/10.1063/1.5111318} {\bibfield  {journal} {\bibinfo
  {journal} {Journal of Applied Physics}\ }\textbf {\bibinfo {volume} {126}},\
  \bibinfo {pages} {085703} (\bibinfo {year} {2019})}\BibitemShut {NoStop}%
\bibitem [{\citenamefont {Bushick}\ \emph {et~al.}(2020)\citenamefont
  {Bushick}, \citenamefont {Mengle}, \citenamefont {Chae},\ and\ \citenamefont
  {Kioupakis}}]{Bushick_APL_2020}%
  \BibitemOpen
  \bibfield  {author} {\bibinfo {author} {\bibfnamefont {K.}~\bibnamefont
  {Bushick}}, \bibinfo {author} {\bibfnamefont {K.~A.}\ \bibnamefont {Mengle}},
  \bibinfo {author} {\bibfnamefont {S.}~\bibnamefont {Chae}},\ and\ \bibinfo
  {author} {\bibfnamefont {E.}~\bibnamefont {Kioupakis}},\ }\bibfield  {title}
  {\bibinfo {title} {Electron and hole mobility of rutile geo2 from first
  principles: An ultrawide-bandgap semiconductor for power electronics},\
  }\href {https://doi.org/10.1063/5.0033284} {\bibfield  {journal} {\bibinfo
  {journal} {Applied Physics Letters}\ }\textbf {\bibinfo {volume} {117}},\
  \bibinfo {pages} {182104} (\bibinfo {year} {2020})}\BibitemShut {NoStop}%
\bibitem [{\citenamefont {Lee}\ \emph {et~al.}(2024)\citenamefont {Lee},
  \citenamefont {Zakutayev},\ and\ \citenamefont
  {Stevanovi{\'c}}}]{Lee_ACEELM_2024}%
  \BibitemOpen
  \bibfield  {author} {\bibinfo {author} {\bibfnamefont {C.-W.}\ \bibnamefont
  {Lee}}, \bibinfo {author} {\bibfnamefont {A.}~\bibnamefont {Zakutayev}},\
  and\ \bibinfo {author} {\bibfnamefont {V.}~\bibnamefont {Stevanovi{\'c}}},\
  }\bibfield  {title} {\bibinfo {title} {Computational insights into phase
  equilibria between wide-gap semiconductors and contact materials},\ }\href
  {https://doi.org/10.1021/acsaelm.4c00032} {\bibfield  {journal} {\bibinfo
  {journal} {ACS Appl. Electron. Mater.}\ }\textbf {\bibinfo {volume} {6}},\
  \bibinfo {pages} {2383} (\bibinfo {year} {2024})}\BibitemShut {NoStop}%
\bibitem [{\citenamefont {Tsao}\ \emph {et~al.}(2018)\citenamefont {Tsao},
  \citenamefont {Chowdhury}, \citenamefont {Hollis}, \citenamefont {Jena},
  \citenamefont {Johnson}, \citenamefont {Jones}, \citenamefont {Kaplar},
  \citenamefont {Rajan}, \citenamefont {Walle}, \citenamefont {Bellotti},
  \citenamefont {Chua}, \citenamefont {Collazo}, \citenamefont {Coltrin},
  \citenamefont {Cooper}, \citenamefont {Evans}, \citenamefont {Graham},
  \citenamefont {Grotjohn}, \citenamefont {Heller}, \citenamefont
  {Higashiwaki}, \citenamefont {Islam}, \citenamefont {Juodawlkis},
  \citenamefont {Khan}, \citenamefont {Koehler}, \citenamefont {Leach},
  \citenamefont {Mishra}, \citenamefont {Nemanich}, \citenamefont
  {Pilawa‐Podgurski}, \citenamefont {Shealy}, \citenamefont {Sitar},
  \citenamefont {Tadjer}, \citenamefont {Witulski}, \citenamefont {Wraback},\
  and\ \citenamefont {Simmons}}]{tsao_ultrawide-bandgap_2018}%
  \BibitemOpen
  \bibfield  {author} {\bibinfo {author} {\bibfnamefont {J.~Y.}\ \bibnamefont
  {Tsao}}, \bibinfo {author} {\bibfnamefont {S.}~\bibnamefont {Chowdhury}},
  \bibinfo {author} {\bibfnamefont {M.~A.}\ \bibnamefont {Hollis}}, \bibinfo
  {author} {\bibfnamefont {D.}~\bibnamefont {Jena}}, \bibinfo {author}
  {\bibfnamefont {N.~M.}\ \bibnamefont {Johnson}}, \bibinfo {author}
  {\bibfnamefont {K.~A.}\ \bibnamefont {Jones}}, \bibinfo {author}
  {\bibfnamefont {R.~J.}\ \bibnamefont {Kaplar}}, \bibinfo {author}
  {\bibfnamefont {S.}~\bibnamefont {Rajan}}, \bibinfo {author} {\bibfnamefont
  {C.~G. V.~d.}\ \bibnamefont {Walle}}, \bibinfo {author} {\bibfnamefont
  {E.}~\bibnamefont {Bellotti}}, \bibinfo {author} {\bibfnamefont {C.~L.}\
  \bibnamefont {Chua}}, \bibinfo {author} {\bibfnamefont {R.}~\bibnamefont
  {Collazo}}, \bibinfo {author} {\bibfnamefont {M.~E.}\ \bibnamefont
  {Coltrin}}, \bibinfo {author} {\bibfnamefont {J.~A.}\ \bibnamefont {Cooper}},
  \bibinfo {author} {\bibfnamefont {K.~R.}\ \bibnamefont {Evans}}, \bibinfo
  {author} {\bibfnamefont {S.}~\bibnamefont {Graham}}, \bibinfo {author}
  {\bibfnamefont {T.~A.}\ \bibnamefont {Grotjohn}}, \bibinfo {author}
  {\bibfnamefont {E.~R.}\ \bibnamefont {Heller}}, \bibinfo {author}
  {\bibfnamefont {M.}~\bibnamefont {Higashiwaki}}, \bibinfo {author}
  {\bibfnamefont {M.~S.}\ \bibnamefont {Islam}}, \bibinfo {author}
  {\bibfnamefont {P.~W.}\ \bibnamefont {Juodawlkis}}, \bibinfo {author}
  {\bibfnamefont {M.~A.}\ \bibnamefont {Khan}}, \bibinfo {author}
  {\bibfnamefont {A.~D.}\ \bibnamefont {Koehler}}, \bibinfo {author}
  {\bibfnamefont {J.~H.}\ \bibnamefont {Leach}}, \bibinfo {author}
  {\bibfnamefont {U.~K.}\ \bibnamefont {Mishra}}, \bibinfo {author}
  {\bibfnamefont {R.~J.}\ \bibnamefont {Nemanich}}, \bibinfo {author}
  {\bibfnamefont {R.~C.~N.}\ \bibnamefont {Pilawa‐Podgurski}}, \bibinfo
  {author} {\bibfnamefont {J.~B.}\ \bibnamefont {Shealy}}, \bibinfo {author}
  {\bibfnamefont {Z.}~\bibnamefont {Sitar}}, \bibinfo {author} {\bibfnamefont
  {M.~J.}\ \bibnamefont {Tadjer}}, \bibinfo {author} {\bibfnamefont {A.~F.}\
  \bibnamefont {Witulski}}, \bibinfo {author} {\bibfnamefont {M.}~\bibnamefont
  {Wraback}},\ and\ \bibinfo {author} {\bibfnamefont {J.~A.}\ \bibnamefont
  {Simmons}},\ }\bibfield  {title} {\bibinfo {title} {Ultrawide-{Bandgap}
  {Semiconductors}: {Research} {Opportunities} and {Challenges}},\ }\href@noop
  {} {\bibfield  {journal} {\bibinfo  {journal} {Advanced Electronic
  Materials}\ }\textbf {\bibinfo {volume} {4}},\ \bibinfo {pages} {1600501}
  (\bibinfo {year} {2018})}\BibitemShut {NoStop}%
\bibitem [{\citenamefont {Yan}\ and\ \citenamefont
  {Wei}(2008)}]{yan_doping_2008}%
  \BibitemOpen
  \bibfield  {author} {\bibinfo {author} {\bibfnamefont {Y.}~\bibnamefont
  {Yan}}\ and\ \bibinfo {author} {\bibfnamefont {S.-H.}\ \bibnamefont {Wei}},\
  }\bibfield  {title} {\bibinfo {title} {Doping asymmetry in wide-bandgap
  semiconductors: {{Origins}} and solutions},\ }\href
  {https://doi.org/10.1002/pssb.200743334} {\bibfield  {journal} {\bibinfo
  {journal} {physica status solidi (b)}\ }\textbf {\bibinfo {volume} {245}},\
  \bibinfo {pages} {641} (\bibinfo {year} {2008})}\BibitemShut {NoStop}%
\bibitem [{\citenamefont {Goyal}\ \emph {et~al.}(2020)\citenamefont {Goyal},
  \citenamefont {Gorai}, \citenamefont {Anand}, \citenamefont {Toberer},
  \citenamefont {Snyder},\ and\ \citenamefont
  {Stevanovi{\'c}}}]{Goyal_CM:2020}%
  \BibitemOpen
  \bibfield  {author} {\bibinfo {author} {\bibfnamefont {A.}~\bibnamefont
  {Goyal}}, \bibinfo {author} {\bibfnamefont {P.}~\bibnamefont {Gorai}},
  \bibinfo {author} {\bibfnamefont {S.}~\bibnamefont {Anand}}, \bibinfo
  {author} {\bibfnamefont {E.~S.}\ \bibnamefont {Toberer}}, \bibinfo {author}
  {\bibfnamefont {G.~J.}\ \bibnamefont {Snyder}},\ and\ \bibinfo {author}
  {\bibfnamefont {V.}~\bibnamefont {Stevanovi{\'c}}},\ }\bibfield  {title}
  {\bibinfo {title} {On the dopability of semiconductors and governing material
  properties},\ }\href {https://doi.org/10.1021/acs.chemmater.9b05126}
  {\bibfield  {journal} {\bibinfo  {journal} {Chemistry of Materials}\ }\textbf
  {\bibinfo {volume} {32}},\ \bibinfo {pages} {4467} (\bibinfo {year}
  {2020})}\BibitemShut {NoStop}%
\bibitem [{\citenamefont {Egbo}\ \emph {et~al.}(2024)\citenamefont {Egbo},
  \citenamefont {Garrity}, \citenamefont {Callahan}, \citenamefont {Chae},
  \citenamefont {Lee}, \citenamefont {Tellekamp}, \citenamefont {Hwang},
  \citenamefont {Stevanovic},\ and\ \citenamefont {Zakutayev}}]{Egbo_APL_2024}%
  \BibitemOpen
  \bibfield  {author} {\bibinfo {author} {\bibfnamefont {K.}~\bibnamefont
  {Egbo}}, \bibinfo {author} {\bibfnamefont {E.~M.}\ \bibnamefont {Garrity}},
  \bibinfo {author} {\bibfnamefont {W.~A.}\ \bibnamefont {Callahan}}, \bibinfo
  {author} {\bibfnamefont {C.}~\bibnamefont {Chae}}, \bibinfo {author}
  {\bibfnamefont {C.-W.}\ \bibnamefont {Lee}}, \bibinfo {author} {\bibfnamefont
  {B.}~\bibnamefont {Tellekamp}}, \bibinfo {author} {\bibfnamefont
  {J.}~\bibnamefont {Hwang}}, \bibinfo {author} {\bibfnamefont
  {V.}~\bibnamefont {Stevanovic}},\ and\ \bibinfo {author} {\bibfnamefont
  {A.}~\bibnamefont {Zakutayev}},\ }\bibfield  {title} {\bibinfo {title}
  {Niga2o4 interfacial layers in nio/ga2o3 heterojunction diodes at high
  temperature},\ }\href {https://doi.org/10.1063/5.0194540} {\bibfield
  {journal} {\bibinfo  {journal} {Applied Physics Letters}\ }\textbf {\bibinfo
  {volume} {124}},\ \bibinfo {pages} {173512} (\bibinfo {year}
  {2024})}\BibitemShut {NoStop}%
\bibitem [{\citenamefont {Gaewdang}\ \emph {et~al.}(1994)\citenamefont
  {Gaewdang}, \citenamefont {Chaminade}, \citenamefont {Gravereau},
  \citenamefont {Garcia}, \citenamefont {Fouassier}, \citenamefont {Pouchard},
  \citenamefont {Hagenmuller},\ and\ \citenamefont
  {Jacquier}}]{gaewdang_structural_1994}%
  \BibitemOpen
  \bibfield  {author} {\bibinfo {author} {\bibfnamefont {T.}~\bibnamefont
  {Gaewdang}}, \bibinfo {author} {\bibfnamefont {J.~P.}\ \bibnamefont
  {Chaminade}}, \bibinfo {author} {\bibfnamefont {P.}~\bibnamefont
  {Gravereau}}, \bibinfo {author} {\bibfnamefont {A.}~\bibnamefont {Garcia}},
  \bibinfo {author} {\bibfnamefont {C.}~\bibnamefont {Fouassier}}, \bibinfo
  {author} {\bibfnamefont {M.}~\bibnamefont {Pouchard}}, \bibinfo {author}
  {\bibfnamefont {P.}~\bibnamefont {Hagenmuller}},\ and\ \bibinfo {author}
  {\bibfnamefont {B.}~\bibnamefont {Jacquier}},\ }\bibfield  {title} {\bibinfo
  {title} {Structural investigations and luminescence of {{In2Ge2O7}} and
  {{In2Si2O7}}},\ }\href {https://doi.org/10.1002/zaac.19946201121} {\bibfield
  {journal} {\bibinfo  {journal} {Zeitschrift f{\"u}r anorganische und
  allgemeine Chemie}\ }\textbf {\bibinfo {volume} {620}},\ \bibinfo {pages}
  {1965} (\bibinfo {year} {1994})}\BibitemShut {NoStop}%
\bibitem [{\citenamefont {Li}\ \emph {et~al.}(2015)\citenamefont {Li},
  \citenamefont {Li}, \citenamefont {Li}, \citenamefont {Zhao}, \citenamefont
  {Zhu}, \citenamefont {Zhu},\ and\ \citenamefont
  {Wang}}]{li_comparative_2015}%
  \BibitemOpen
  \bibfield  {author} {\bibinfo {author} {\bibfnamefont {H.}~\bibnamefont
  {Li}}, \bibinfo {author} {\bibfnamefont {Y.}~\bibnamefont {Li}}, \bibinfo
  {author} {\bibfnamefont {N.}~\bibnamefont {Li}}, \bibinfo {author}
  {\bibfnamefont {Y.}~\bibnamefont {Zhao}}, \bibinfo {author} {\bibfnamefont
  {H.}~\bibnamefont {Zhu}}, \bibinfo {author} {\bibfnamefont {P.}~\bibnamefont
  {Zhu}},\ and\ \bibinfo {author} {\bibfnamefont {X.}~\bibnamefont {Wang}},\
  }\bibfield  {title} {\bibinfo {title} {A comparative study of high pressure
  behaviors of pyrochlore-type and thortveitite-type {{In2Ge2O7}}},\ }\href
  {https://doi.org/10.1039/C5RA04587H} {\bibfield  {journal} {\bibinfo
  {journal} {RSC Advances}\ }\textbf {\bibinfo {volume} {5}},\ \bibinfo {pages}
  {44121} (\bibinfo {year} {2015})}\BibitemShut {NoStop}%
\bibitem [{\citenamefont {Shannon}\ and\ \citenamefont
  {Prewitt}(1970)}]{shannon_synthesis_1970}%
  \BibitemOpen
  \bibfield  {author} {\bibinfo {author} {\bibfnamefont {R.}~\bibnamefont
  {Shannon}}\ and\ \bibinfo {author} {\bibfnamefont {C.}~\bibnamefont
  {Prewitt}},\ }\bibfield  {title} {\bibinfo {title} {Synthesis of
  pyrosilicates and pyrogermanates having the thortveitite structure},\ }\href
  {https://doi.org/10.1016/0022-4596(70)90070-8} {\bibfield  {journal}
  {\bibinfo  {journal} {Journal of Solid State Chemistry}\ }\textbf {\bibinfo
  {volume} {2}},\ \bibinfo {pages} {199} (\bibinfo {year} {1970})}\BibitemShut
  {NoStop}%
\bibitem [{\citenamefont {Reid}\ \emph {et~al.}(1977)\citenamefont {Reid},
  \citenamefont {Li},\ and\ \citenamefont
  {Ringwood}}]{reid_high-pressure_1977}%
  \BibitemOpen
  \bibfield  {author} {\bibinfo {author} {\bibfnamefont {A.~F.}\ \bibnamefont
  {Reid}}, \bibinfo {author} {\bibfnamefont {C.}~\bibnamefont {Li}},\ and\
  \bibinfo {author} {\bibfnamefont {A.~E.}\ \bibnamefont {Ringwood}},\
  }\bibfield  {title} {\bibinfo {title} {High-pressure silicate pyrochlores,
  {{Sc2Si2O7}} and {{In2Si2O7}}},\ }\href
  {https://doi.org/10.1016/0022-4596(77)90157-8} {\bibfield  {journal}
  {\bibinfo  {journal} {Journal of Solid State Chemistry}\ }\textbf {\bibinfo
  {volume} {20}},\ \bibinfo {pages} {219} (\bibinfo {year} {1977})}\BibitemShut
  {NoStop}%
\bibitem [{\citenamefont {Shannon}\ and\ \citenamefont
  {Sleight}(1968)}]{shannon_synthesis_1968}%
  \BibitemOpen
  \bibfield  {author} {\bibinfo {author} {\bibfnamefont {R.~D.}\ \bibnamefont
  {Shannon}}\ and\ \bibinfo {author} {\bibfnamefont {A.~W.}\ \bibnamefont
  {Sleight}},\ }\bibfield  {title} {\bibinfo {title} {Synthesis of new
  high-pressure pyrochlore phases},\ }\href
  {https://doi.org/10.1021/ic50066a038} {\bibfield  {journal} {\bibinfo
  {journal} {Inorganic Chemistry}\ }\textbf {\bibinfo {volume} {7}},\ \bibinfo
  {pages} {1649} (\bibinfo {year} {1968})}\BibitemShut {NoStop}%
\bibitem [{\citenamefont {Messous}\ \emph {et~al.}(1995)\citenamefont
  {Messous}, \citenamefont {Chambon}, \citenamefont {De~J{\'e}sus},
  \citenamefont {Drain}, \citenamefont {Pastor}, \citenamefont {Garcia},
  \citenamefont {Chaminade}, \citenamefont {Gaewdang}, \citenamefont
  {Fouassier}, \citenamefont {Jacquier},\ and\ \citenamefont
  {Varrel}}]{messous_indium_1995}%
  \BibitemOpen
  \bibfield  {author} {\bibinfo {author} {\bibfnamefont {Y.}~\bibnamefont
  {Messous}}, \bibinfo {author} {\bibfnamefont {B.}~\bibnamefont {Chambon}},
  \bibinfo {author} {\bibfnamefont {M.}~\bibnamefont {De~J{\'e}sus}}, \bibinfo
  {author} {\bibfnamefont {D.}~\bibnamefont {Drain}}, \bibinfo {author}
  {\bibfnamefont {C.}~\bibnamefont {Pastor}}, \bibinfo {author} {\bibfnamefont
  {A.}~\bibnamefont {Garcia}}, \bibinfo {author} {\bibfnamefont
  {J.}~\bibnamefont {Chaminade}}, \bibinfo {author} {\bibfnamefont
  {T.}~\bibnamefont {Gaewdang}}, \bibinfo {author} {\bibfnamefont
  {C.}~\bibnamefont {Fouassier}}, \bibinfo {author} {\bibfnamefont
  {B.}~\bibnamefont {Jacquier}},\ and\ \bibinfo {author} {\bibfnamefont
  {B.}~\bibnamefont {Varrel}},\ }\bibfield  {title} {\bibinfo {title} {Indium
  disilicate, a new fast scintillator},\ }\href
  {https://doi.org/10.1016/0168-9002(94)01007-2} {\bibfield  {journal}
  {\bibinfo  {journal} {Nuclear Instruments and Methods in Physics Research
  Section A: Accelerators, Spectrometers, Detectors and Associated Equipment}\
  }\textbf {\bibinfo {volume} {354}},\ \bibinfo {pages} {527} (\bibinfo {year}
  {1995})}\BibitemShut {NoStop}%
\bibitem [{\citenamefont {Stevanovi\ifmmode~\acute{c}\else \'{c}\fi{}}\ \emph
  {et~al.}(2012)\citenamefont {Stevanovi\ifmmode~\acute{c}\else \'{c}\fi{}},
  \citenamefont {Lany}, \citenamefont {Zhang},\ and\ \citenamefont
  {Zunger}}]{Stevanovic_PRB:2012}%
  \BibitemOpen
  \bibfield  {author} {\bibinfo {author} {\bibfnamefont {V.}~\bibnamefont
  {Stevanovi\ifmmode~\acute{c}\else \'{c}\fi{}}}, \bibinfo {author}
  {\bibfnamefont {S.}~\bibnamefont {Lany}}, \bibinfo {author} {\bibfnamefont
  {X.}~\bibnamefont {Zhang}},\ and\ \bibinfo {author} {\bibfnamefont
  {A.}~\bibnamefont {Zunger}},\ }\bibfield  {title} {\bibinfo {title}
  {Correcting density functional theory for accurate predictions of compound
  enthalpies of formation: Fitted elemental-phase reference energies},\ }\href
  {https://doi.org/10.1103/PhysRevB.85.115104} {\bibfield  {journal} {\bibinfo
  {journal} {Phys. Rev. B}\ }\textbf {\bibinfo {volume} {85}},\ \bibinfo
  {pages} {115104} (\bibinfo {year} {2012})}\BibitemShut {NoStop}%
\bibitem [{\citenamefont {Perdew}\ \emph {et~al.}(1996)\citenamefont {Perdew},
  \citenamefont {Burke},\ and\ \citenamefont {Ernzerhof}}]{PBE_PRL:1996}%
  \BibitemOpen
  \bibfield  {author} {\bibinfo {author} {\bibfnamefont {J.~P.}\ \bibnamefont
  {Perdew}}, \bibinfo {author} {\bibfnamefont {K.}~\bibnamefont {Burke}},\ and\
  \bibinfo {author} {\bibfnamefont {M.}~\bibnamefont {Ernzerhof}},\ }\bibfield
  {title} {\bibinfo {title} {Generalized gradient approximation made simple},\
  }\href {https://doi.org/10.1103/PhysRevLett.77.3865} {\bibfield  {journal}
  {\bibinfo  {journal} {Phys. Rev. Lett.}\ }\textbf {\bibinfo {volume} {77}},\
  \bibinfo {pages} {3865} (\bibinfo {year} {1996})}\BibitemShut {NoStop}%
\bibitem [{\citenamefont {Bl\"ochl}(1994)}]{PAW_PRB:1994}%
  \BibitemOpen
  \bibfield  {author} {\bibinfo {author} {\bibfnamefont {P.~E.}\ \bibnamefont
  {Bl\"ochl}},\ }\bibfield  {title} {\bibinfo {title} {Projector augmented-wave
  method},\ }\href {https://doi.org/10.1103/PhysRevB.50.17953} {\bibfield
  {journal} {\bibinfo  {journal} {Phys. Rev. B}\ }\textbf {\bibinfo {volume}
  {50}},\ \bibinfo {pages} {17953} (\bibinfo {year} {1994})}\BibitemShut
  {NoStop}%
\bibitem [{\citenamefont {Kresse}\ and\ \citenamefont
  {Joubert}(1999)}]{Kresse_PRB_1999}%
  \BibitemOpen
  \bibfield  {author} {\bibinfo {author} {\bibfnamefont {G.}~\bibnamefont
  {Kresse}}\ and\ \bibinfo {author} {\bibfnamefont {D.}~\bibnamefont
  {Joubert}},\ }\bibfield  {title} {\bibinfo {title} {From ultrasoft
  pseudopotentials to the projector augmented-wave method},\ }\href
  {https://doi.org/10.1103/PhysRevB.59.1758} {\bibfield  {journal} {\bibinfo
  {journal} {Phys. Rev. B}\ }\textbf {\bibinfo {volume} {59}},\ \bibinfo
  {pages} {1758} (\bibinfo {year} {1999})}\BibitemShut {NoStop}%
\bibitem [{\citenamefont {Zagorac}\ \emph {et~al.}(2019)\citenamefont
  {Zagorac}, \citenamefont {M{\"{u}}ller}, \citenamefont {Ruehl}, \citenamefont
  {Zagorac},\ and\ \citenamefont {Rehme}}]{icsd}%
  \BibitemOpen
  \bibfield  {author} {\bibinfo {author} {\bibfnamefont {D.}~\bibnamefont
  {Zagorac}}, \bibinfo {author} {\bibfnamefont {H.}~\bibnamefont
  {M{\"{u}}ller}}, \bibinfo {author} {\bibfnamefont {S.}~\bibnamefont {Ruehl}},
  \bibinfo {author} {\bibfnamefont {J.}~\bibnamefont {Zagorac}},\ and\ \bibinfo
  {author} {\bibfnamefont {S.}~\bibnamefont {Rehme}},\ }\bibfield  {title}
  {\bibinfo {title} {{Recent developments in the Inorganic Crystal Structure
  Database: theoretical crystal structure data and related features}},\ }\href
  {https://doi.org/10.1107/S160057671900997X} {\bibfield  {journal} {\bibinfo
  {journal} {Journal of Applied Crystallography}\ }\textbf {\bibinfo {volume}
  {52}},\ \bibinfo {pages} {918} (\bibinfo {year} {2019})}\BibitemShut
  {NoStop}%
\bibitem [{\citenamefont {Birch}(1947)}]{Birch_PR:1947}%
  \BibitemOpen
  \bibfield  {author} {\bibinfo {author} {\bibfnamefont {F.}~\bibnamefont
  {Birch}},\ }\bibfield  {title} {\bibinfo {title} {Finite elastic strain of
  cubic crystals},\ }\href {https://doi.org/10.1103/PhysRev.71.809} {\bibfield
  {journal} {\bibinfo  {journal} {Phys. Rev.}\ }\textbf {\bibinfo {volume}
  {71}},\ \bibinfo {pages} {809} (\bibinfo {year} {1947})}\BibitemShut
  {NoStop}%
\bibitem [{\citenamefont {Sheppard}\ \emph {et~al.}(2012)\citenamefont
  {Sheppard}, \citenamefont {Xiao}, \citenamefont {Chemelewski}, \citenamefont
  {Johnson},\ and\ \citenamefont {Henkelman}}]{ssneb}%
  \BibitemOpen
  \bibfield  {author} {\bibinfo {author} {\bibfnamefont {D.}~\bibnamefont
  {Sheppard}}, \bibinfo {author} {\bibfnamefont {P.}~\bibnamefont {Xiao}},
  \bibinfo {author} {\bibfnamefont {W.}~\bibnamefont {Chemelewski}}, \bibinfo
  {author} {\bibfnamefont {D.~D.}\ \bibnamefont {Johnson}},\ and\ \bibinfo
  {author} {\bibfnamefont {G.}~\bibnamefont {Henkelman}},\ }\bibfield  {title}
  {\bibinfo {title} {A generalized solid-state nudged elastic band method},\
  }\href {https://doi.org/10.1063/1.3684549} {\bibfield  {journal} {\bibinfo
  {journal} {The Journal of Chemical Physics}\ }\textbf {\bibinfo {volume}
  {136}},\ \bibinfo {pages} {074103} (\bibinfo {year} {2012})},\ \Eprint
  {https://arxiv.org/abs/https://pubs.aip.org/aip/jcp/article-pdf/doi/10.1063/1.3684549/13348476/074103\_1\_online.pdf}
  {https://pubs.aip.org/aip/jcp/article-pdf/doi/10.1063/1.3684549/13348476/074103\_1\_online.pdf}
  \BibitemShut {NoStop}%
\bibitem [{\citenamefont {Stevanovi\ifmmode~\acute{c}\else \'{c}\fi{}}\ \emph
  {et~al.}(2018)\citenamefont {Stevanovi\ifmmode~\acute{c}\else \'{c}\fi{}},
  \citenamefont {Trottier}, \citenamefont {Musgrave}, \citenamefont {Therrien},
  \citenamefont {Holder},\ and\ \citenamefont {Graf}}]{Stevanovic_PRM:2018}%
  \BibitemOpen
  \bibfield  {author} {\bibinfo {author} {\bibfnamefont {V.}~\bibnamefont
  {Stevanovi\ifmmode~\acute{c}\else \'{c}\fi{}}}, \bibinfo {author}
  {\bibfnamefont {R.}~\bibnamefont {Trottier}}, \bibinfo {author}
  {\bibfnamefont {C.}~\bibnamefont {Musgrave}}, \bibinfo {author}
  {\bibfnamefont {F.}~\bibnamefont {Therrien}}, \bibinfo {author}
  {\bibfnamefont {A.}~\bibnamefont {Holder}},\ and\ \bibinfo {author}
  {\bibfnamefont {P.}~\bibnamefont {Graf}},\ }\bibfield  {title} {\bibinfo
  {title} {Predicting kinetics of polymorphic transformations from structure
  mapping and coordination analysis},\ }\href
  {https://doi.org/10.1103/PhysRevMaterials.2.033802} {\bibfield  {journal}
  {\bibinfo  {journal} {Phys. Rev. Mater.}\ }\textbf {\bibinfo {volume} {2}},\
  \bibinfo {pages} {033802} (\bibinfo {year} {2018})}\BibitemShut {NoStop}%
\bibitem [{\citenamefont {Therrien}\ \emph {et~al.}(2020)\citenamefont
  {Therrien}, \citenamefont {Graf},\ and\ \citenamefont
  {Stevanović}}]{Therrien_JCP:2020}%
  \BibitemOpen
  \bibfield  {author} {\bibinfo {author} {\bibfnamefont {F.}~\bibnamefont
  {Therrien}}, \bibinfo {author} {\bibfnamefont {P.}~\bibnamefont {Graf}},\
  and\ \bibinfo {author} {\bibfnamefont {V.}~\bibnamefont {Stevanović}},\
  }\bibfield  {title} {\bibinfo {title} {Matching crystal structures
  atom-to-atom},\ }\href {https://doi.org/10.1063/1.5131527} {\bibfield
  {journal} {\bibinfo  {journal} {The Journal of Chemical Physics}\ }\textbf
  {\bibinfo {volume} {152}},\ \bibinfo {pages} {074106} (\bibinfo {year}
  {2020})},\ \Eprint
  {https://arxiv.org/abs/https://pubs.aip.org/aip/jcp/article-pdf/doi/10.1063/1.5131527/15570170/074106\_1\_online.pdf}
  {https://pubs.aip.org/aip/jcp/article-pdf/doi/10.1063/1.5131527/15570170/074106\_1\_online.pdf}
  \BibitemShut {NoStop}%
\bibitem [{\citenamefont {Peng}\ \emph {et~al.}(2013)\citenamefont {Peng},
  \citenamefont {Scanlon}, \citenamefont {Stevanovic}, \citenamefont {Vidal},
  \citenamefont {Watson},\ and\ \citenamefont {Lany}}]{Peng_PRB_2013}%
  \BibitemOpen
  \bibfield  {author} {\bibinfo {author} {\bibfnamefont {H.}~\bibnamefont
  {Peng}}, \bibinfo {author} {\bibfnamefont {D.~O.}\ \bibnamefont {Scanlon}},
  \bibinfo {author} {\bibfnamefont {V.}~\bibnamefont {Stevanovic}}, \bibinfo
  {author} {\bibfnamefont {J.}~\bibnamefont {Vidal}}, \bibinfo {author}
  {\bibfnamefont {G.~W.}\ \bibnamefont {Watson}},\ and\ \bibinfo {author}
  {\bibfnamefont {S.}~\bibnamefont {Lany}},\ }\bibfield  {title} {\bibinfo
  {title} {Convergence of density and hybrid functional defect calculations for
  compound semiconductors},\ }\href
  {https://doi.org/10.1103/PhysRevB.88.115201} {\bibfield  {journal} {\bibinfo
  {journal} {Phys. Rev. B}\ }\textbf {\bibinfo {volume} {88}},\ \bibinfo
  {pages} {115201} (\bibinfo {year} {2013})}\BibitemShut {NoStop}%
\bibitem [{\citenamefont {Lany}(2013)}]{Lany_PRB_2013}%
  \BibitemOpen
  \bibfield  {author} {\bibinfo {author} {\bibfnamefont {S.}~\bibnamefont
  {Lany}},\ }\bibfield  {title} {\bibinfo {title} {Band-structure calculations
  for the 3$d$ transition metal oxides in $gw$},\ }\href
  {https://doi.org/10.1103/PhysRevB.87.085112} {\bibfield  {journal} {\bibinfo
  {journal} {Phys. Rev. B}\ }\textbf {\bibinfo {volume} {87}},\ \bibinfo
  {pages} {085112} (\bibinfo {year} {2013})}\BibitemShut {NoStop}%
\bibitem [{\citenamefont {Yan}\ \emph {et~al.}(2015)\citenamefont {Yan},
  \citenamefont {Gorai}, \citenamefont {Ortiz}, \citenamefont {Miller},
  \citenamefont {Barnett}, \citenamefont {Mason}, \citenamefont {Stevanović},\
  and\ \citenamefont {Toberer}}]{Yan_EES_2015}%
  \BibitemOpen
  \bibfield  {author} {\bibinfo {author} {\bibfnamefont {J.}~\bibnamefont
  {Yan}}, \bibinfo {author} {\bibfnamefont {P.}~\bibnamefont {Gorai}}, \bibinfo
  {author} {\bibfnamefont {B.}~\bibnamefont {Ortiz}}, \bibinfo {author}
  {\bibfnamefont {S.}~\bibnamefont {Miller}}, \bibinfo {author} {\bibfnamefont
  {S.~A.}\ \bibnamefont {Barnett}}, \bibinfo {author} {\bibfnamefont
  {T.}~\bibnamefont {Mason}}, \bibinfo {author} {\bibfnamefont
  {V.}~\bibnamefont {Stevanović}},\ and\ \bibinfo {author} {\bibfnamefont
  {E.~S.}\ \bibnamefont {Toberer}},\ }\bibfield  {title} {\bibinfo {title}
  {Material descriptors for predicting thermoelectric performance},\
  }\href@noop {} {\bibfield  {journal} {\bibinfo  {journal} {Energy \&
  Environmental Science}\ }\textbf {\bibinfo {volume} {8}},\ \bibinfo {pages}
  {983} (\bibinfo {year} {2015})}\BibitemShut {NoStop}%
\bibitem [{\citenamefont {Lany}\ and\ \citenamefont
  {Zunger}(2009)}]{Lany_MSMSE_2009}%
  \BibitemOpen
  \bibfield  {author} {\bibinfo {author} {\bibfnamefont {S.}~\bibnamefont
  {Lany}}\ and\ \bibinfo {author} {\bibfnamefont {A.}~\bibnamefont {Zunger}},\
  }\bibfield  {title} {\bibinfo {title} {Accurate prediction of defect
  properties in density functional supercell calculations},\ }\href
  {https://doi.org/10.1088/0965-0393/17/8/084002} {\bibfield  {journal}
  {\bibinfo  {journal} {Modelling and Simulation in Materials Science and
  Engineering}\ }\textbf {\bibinfo {volume} {17}},\ \bibinfo {pages} {084002}
  (\bibinfo {year} {2009})}\BibitemShut {NoStop}%
\bibitem [{\citenamefont {Goyal}\ \emph {et~al.}(2017)\citenamefont {Goyal},
  \citenamefont {Gorai}, \citenamefont {Peng}, \citenamefont {Lany},\ and\
  \citenamefont {Stevanović}}]{Goyal_CMS_2017}%
  \BibitemOpen
  \bibfield  {author} {\bibinfo {author} {\bibfnamefont {A.}~\bibnamefont
  {Goyal}}, \bibinfo {author} {\bibfnamefont {P.}~\bibnamefont {Gorai}},
  \bibinfo {author} {\bibfnamefont {H.}~\bibnamefont {Peng}}, \bibinfo {author}
  {\bibfnamefont {S.}~\bibnamefont {Lany}},\ and\ \bibinfo {author}
  {\bibfnamefont {V.}~\bibnamefont {Stevanović}},\ }\bibfield  {title}
  {\bibinfo {title} {A computational framework for automation of point defect
  calculations},\ }\href
  {https://doi.org/https://doi.org/10.1016/j.commatsci.2016.12.040} {\bibfield
  {journal} {\bibinfo  {journal} {Computational Materials Science}\ }\textbf
  {\bibinfo {volume} {130}},\ \bibinfo {pages} {1 } (\bibinfo {year}
  {2017})}\BibitemShut {NoStop}%
\bibitem [{\citenamefont {Goyal}\ and\ \citenamefont
  {Stevanovi\ifmmode~\acute{c}\else \'{c}\fi{}}(2018)}]{Goyal_PRMater_2018}%
  \BibitemOpen
  \bibfield  {author} {\bibinfo {author} {\bibfnamefont {A.}~\bibnamefont
  {Goyal}}\ and\ \bibinfo {author} {\bibfnamefont {V.}~\bibnamefont
  {Stevanovi\ifmmode~\acute{c}\else \'{c}\fi{}}},\ }\bibfield  {title}
  {\bibinfo {title} {Metastable rocksalt zno is $p$-type dopable},\ }\href
  {https://doi.org/10.1103/PhysRevMaterials.2.084603} {\bibfield  {journal}
  {\bibinfo  {journal} {Phys. Rev. Mater.}\ }\textbf {\bibinfo {volume} {2}},\
  \bibinfo {pages} {084603} (\bibinfo {year} {2018})}\BibitemShut {NoStop}%
\bibitem [{\citenamefont {Lany}(2018)}]{Lany_JCP_2018}%
  \BibitemOpen
  \bibfield  {author} {\bibinfo {author} {\bibfnamefont {S.}~\bibnamefont
  {Lany}},\ }\bibfield  {title} {\bibinfo {title} {Communication: The
  electronic entropy of charged defect formation and its impact on
  thermochemical redox cycles},\ }\href {https://doi.org/10.1063/1.5022176}
  {\bibfield  {journal} {\bibinfo  {journal} {The Journal of Chemical Physics}\
  }\textbf {\bibinfo {volume} {148}},\ \bibinfo {pages} {071101} (\bibinfo
  {year} {2018})}\BibitemShut {NoStop}%
\bibitem [{\citenamefont {Callahan}\ \emph {et~al.}(2024)\citenamefont
  {Callahan}, \citenamefont {Egbo}, \citenamefont {Lee}, \citenamefont
  {Ginley}, \citenamefont {O'Hayre},\ and\ \citenamefont
  {Zakutayev}}]{Callahan_APL_2024}%
  \BibitemOpen
  \bibfield  {author} {\bibinfo {author} {\bibfnamefont {W.~A.}\ \bibnamefont
  {Callahan}}, \bibinfo {author} {\bibfnamefont {K.}~\bibnamefont {Egbo}},
  \bibinfo {author} {\bibfnamefont {C.-W.}\ \bibnamefont {Lee}}, \bibinfo
  {author} {\bibfnamefont {D.}~\bibnamefont {Ginley}}, \bibinfo {author}
  {\bibfnamefont {R.}~\bibnamefont {O'Hayre}},\ and\ \bibinfo {author}
  {\bibfnamefont {A.}~\bibnamefont {Zakutayev}},\ }\bibfield  {title} {\bibinfo
  {title} {Reliable operation of cr2o3:mg/$\beta$-ga2o3 p–n heterojunction
  diodes at 600$^{\circ}$c},\ }\href {https://doi.org/10.1063/5.0185566}
  {\bibfield  {journal} {\bibinfo  {journal} {Applied Physics Letters}\
  }\textbf {\bibinfo {volume} {124}},\ \bibinfo {pages} {153504} (\bibinfo
  {year} {2024})}\BibitemShut {NoStop}%
\bibitem [{\citenamefont {Bierwagen}\ and\ \citenamefont
  {Speck}(2010)}]{Bierwagen_JAP_2010}%
  \BibitemOpen
  \bibfield  {author} {\bibinfo {author} {\bibfnamefont {O.}~\bibnamefont
  {Bierwagen}}\ and\ \bibinfo {author} {\bibfnamefont {J.~S.}\ \bibnamefont
  {Speck}},\ }\bibfield  {title} {\bibinfo {title} {Nucleation of islands and
  continuous high-quality in2o3(001) films during plasma-assisted molecular
  beam epitaxy on y-stabilized zro2(001)},\ }\href
  {https://doi.org/10.1063/1.3415539} {\bibfield  {journal} {\bibinfo
  {journal} {Journal of Applied Physics}\ }\textbf {\bibinfo {volume} {107}},\
  \bibinfo {pages} {113519} (\bibinfo {year} {2010})}\BibitemShut {NoStop}%
\end{thebibliography}%

\end{document}



\title{--- Supplementary Material ---\\Stability, growth, and doping of In$_{2}$(Si, Ge)$_{2}$O$_{7}$ as promising n-type wide-gap semiconductors}

\author{Cheng-Wei Lee}

\author{Kingsley Egbo}

\author{Emily Garrity}

\author{Matthew Jankousky}

\author{Henry Garland}

\author{Andriy Zakutayev}
\email{andriy.zakutayev@nrel.gov}

\author{Vladan Stevanovi\'c}
\email{vstevano@mines.edu}
\affiliation{Colorado School of Mines, Golden, CO 80401,USA}
\affiliation{National Renewable Energy Laboratory, Golden, CO 80401,USA }

\maketitle

\setcounter{figure}{0}
\setcounter{table}{0}
\renewcommand{\thetable}{S\arabic{table}}
\renewcommand{\thefigure}{S\arabic{figure}}

\begin{figure}[!htb]
\includegraphics[width=0.5\linewidth]{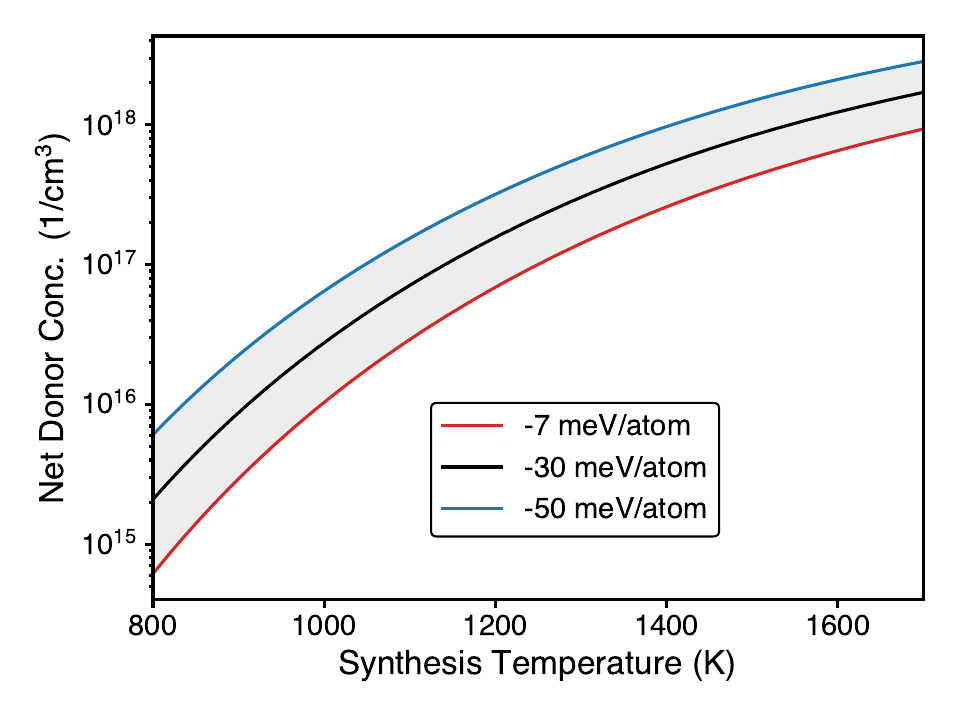}
\caption{\label{fig:fere-correction} Net donor concentration as a function of synthesis temperature for Zr-doped T-In$_2$Si$_2$O$_7$, with different corrections on total energy of T-In$_2$Si$_2$O$_7$. Results with correction of -30 meV/atom were reported in the main text and the difference from results of -7 meV/atom or -50 meV/atom is within one order of magnitude.}
\end{figure}

\begin{figure}[!htb]
\centering
\includegraphics[width=0.5\linewidth]{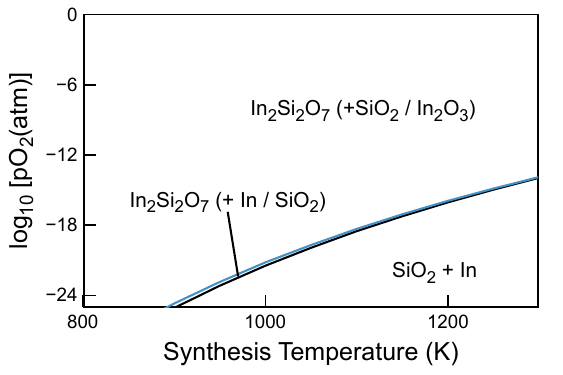}
\caption{\label{fig:pO2-T-Si} 
Predicted partial oxygen pressure (pO$_2$) \textit{vs.} synthesis temperature phase diagrams for T-In$_2$Si$_2$O$_7$. Parentheses indicate phases that can coexist when stoichiometry between In and Ge deviates from 1:1. Blue curves indicate the oxygen-poor condition.
}
\end{figure}

\begin{figure}[!htb]
\includegraphics[width=0.5\linewidth]{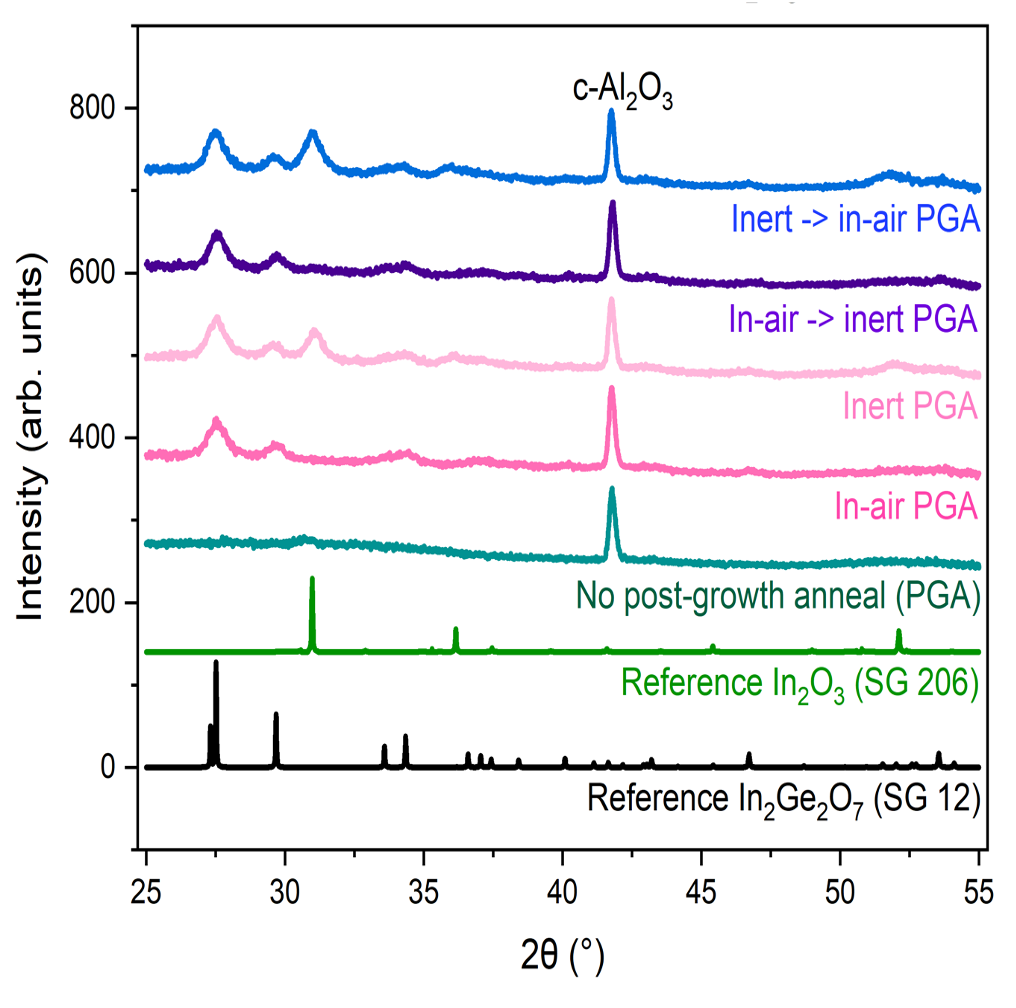}
\caption{\label{fig:xrd-anneal} 2Theta-Omega scans of thin films deposited on $\alpha$-Al$_2$O$_3$(00.1) on the substrates at 600$^{\circ}$C under different annealing environments. PGA indicates post-growth annealing. In$_2$O$_3$ forms if film is annealed in inert atmosphere and initial annealing environment determines whether In$_2$O$_3$ forms.
}
\end{figure}

\begin{figure}
\includegraphics[width=0.95\linewidth]{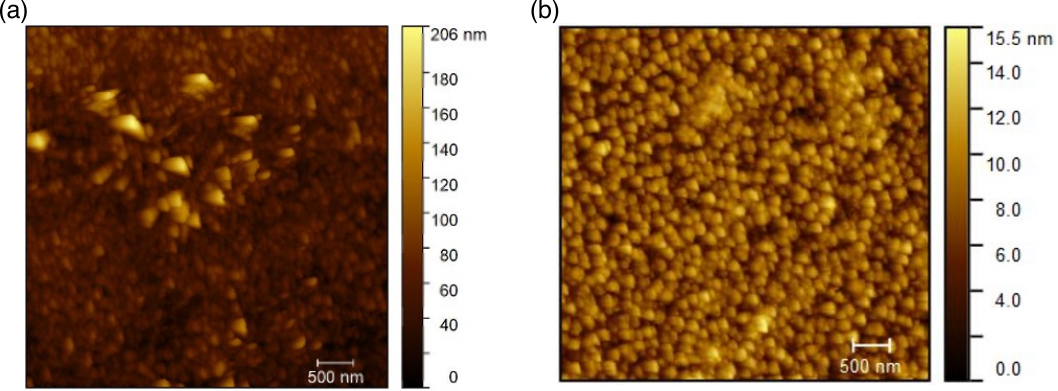}
\caption{\label{fig:AFM} AFM micrographs of T-In$_2$Ge$_2$O$_7$ thin films grown by PLD. (a) Stoichiometric T-In$_2$Ge$_2$O$_7$ thin films with Root-mean-squared roughness $R_q$, of 10.51 nm  (b) In-rich In$_2$Ge$_2$O$_7$ with $R_q$ value of  $\sim$8.1 nm. The size of the images is 5 x 5 $\mu$m$^2$. 
}
\end{figure}

\begin{figure}
\includegraphics[width=0.95\linewidth]{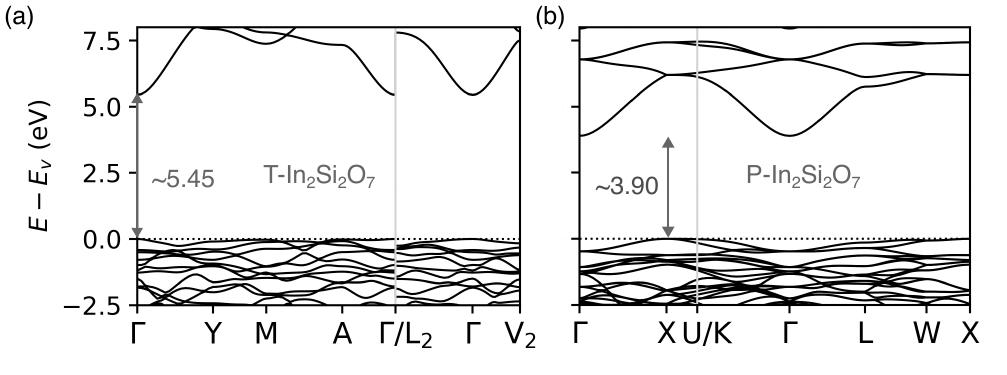}
\caption{\label{fig:es-Si} Predicted electronic structure for both (a) T- and (b) P-In$_2$Si$_2$O$_7$. $E_v$ is the valence band maximum and GW correction was applied as a rigid scissor shift to valence and conduction band edges. T- and P-In$_2$Si$_2$O$_7$ have direct and indirect band gap, respectively.
}
\end{figure}

\begin{figure}[!ht]
\centering
\includegraphics[width=0.9\linewidth]{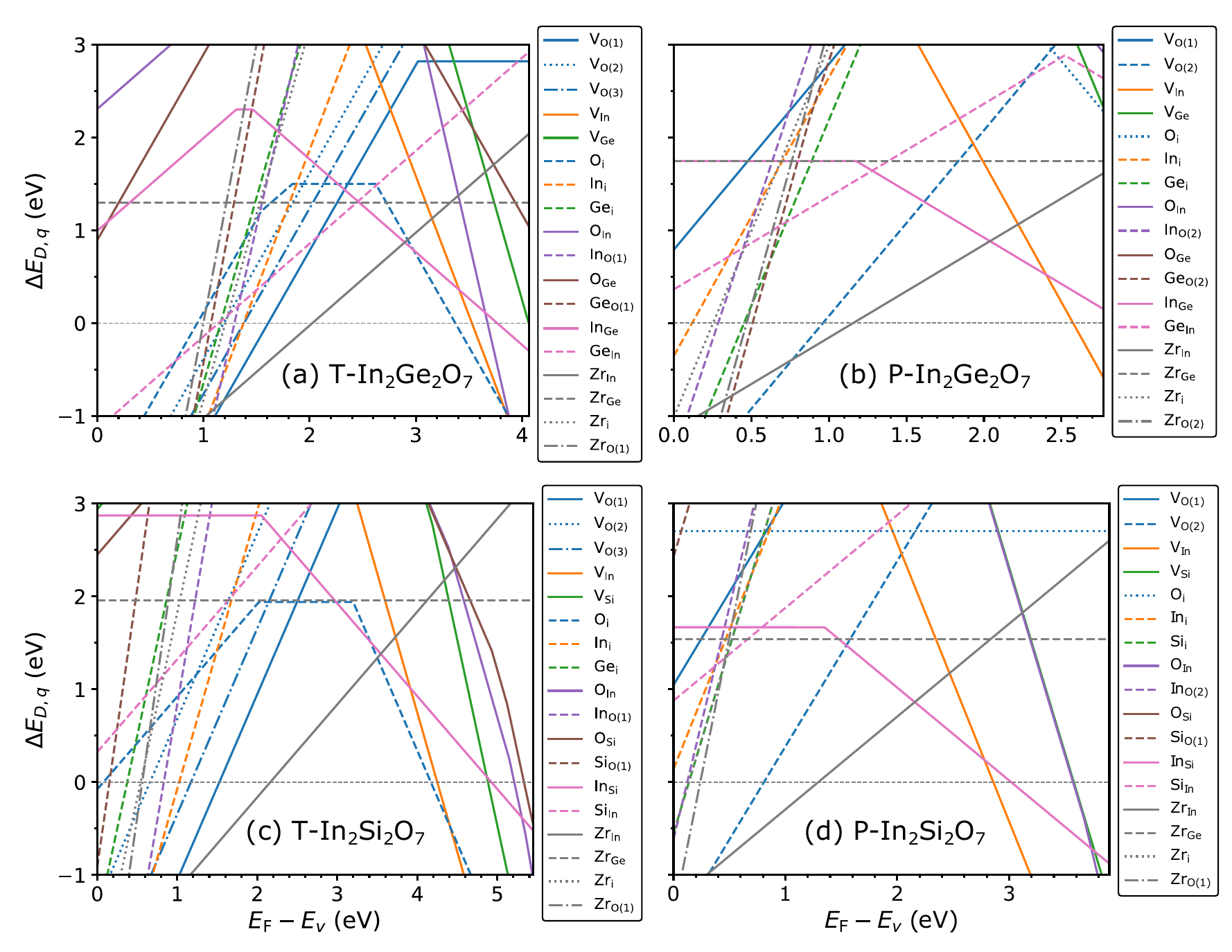}
\caption{\label{figs1} 
Defect formation energies as a function of Fermi energy under O-rich condition (see vertices V3 on Table S4--7) for native defects and Zr impurities in (a) T-In$_{2}$Ge$_{2}$O$_{7}$ under 6 GPa, (b) P-In$_{2}$Ge$_{2}$O$_{7}$, (c) T-In$_{2}$Si$_{2}$O$_{7}$ under 12 GP, and (d) P-In$_{2}$Si$_{2}$O$_{7}$. O(1), O(2), and O(3) indicate different oxygen Wyckoff sites. 
}
\end{figure}

\newpage
\begin{table}[!ht]
\caption{Fitted elemental reference chemical potentials in eV under standard conditions. The reference chemical potentials under pressure (P) of 6 GPa and/or 12 GPa for In, Ge, Si, and Zr are also included. $\mu^0$ (P) =$\mu^0$+ $\Delta H$, where  $\Delta$ H(P) = H(P) - H$^{0}$, as shown in Table II and III.}
\begin{tabular}{cccc}
\hline
& \\
Element & $\mu^0$  & $\mu^0$(6GPa) &$\mu^0$(12GPa)\\
& \\
\hline
& \\
In & -2.310 & -1.343 & -0.468 \\
Ge & -4.140 & -3.284& -\\
Si & -4.990 & - & -2.693\\
O & -4.760 & -& -\\
Zr & -5.870 & -4.954 &-4.084\\
Sn & -3.790 & - & - \\
Hf & -7.400 & - & - \\
Pb & -3.850 & - & - \\
As & -5.060 & - & - \\
V & -6.420 & - & - \\
Bi & -4.390 & - & - \\
& \\
\hline
\end{tabular}\\
\label{table:chempot}
\end{table}

\begin{table}[!ht]
\caption{Enthalpies (H) of all the considered phases under pressures of 12 GPa for In-Si-Zr-O system. H$^o$ and V$_0$ is the total energy and equilibrium volume of the relaxed structure without external pressure.  U and V are the total energy and volume under the given pressure (P) and H=U+PV. 1 GPa $\approx$ 0.0062415 eV/\AA$^3$ was used. Correction to T-In$_2$Si$_2$O$_7$ is not included.} 
\begin{tabular}{cccccc}
\hline
& \\
material & H$_0$ (eV/atom) & V$_0$ (\AA$^{3}$/atom) & H (eV/atom) & U (eV/atom) & V (\AA$^{3}$/atom) \\
& \\
\hline 
& \\
Si & -5.424 & 20.444 & -3.977 & -5.353 & 18.366 \vspace{0.2 cm} \\ 
In & -2.726 & 27.620 & -0.884 & -2.567 & 22.469 \vspace{0.2 cm} \\ 
Zr & -6.598 & 25.235 & -4.812 & -6.511 & 22.689 \vspace{0.2 cm} \\ 
rutile SiO$_2$  & -7.729 & 8.060 & -7.136 & -7.718 & 7.767 \vspace{0.2 cm} \\ 
ZrSiO$_4$ & -8.468 & 11.523 & -7.626 & -8.448 & 10.969 \vspace{0.2 cm} \\ 
P-In$_2$Si$_2$O$_7$ & -6.773 & 9.905 & -6.049 & -6.757 & 9.449 \vspace{0.2 cm} \\ 
T-In$_2$Si$_2$O$_7$ & -6.888 & 12.470 & -5.982 & -6.862 & 11.755 \vspace{0.2 cm} \\
In$_2$O$_3$ & -5.676 & 13.683 & -4.688 & -5.643 & 12.750 \vspace{0.2 cm} \\ 
ZrO$_2$ & -8.964 & 12.586 & -8.055 & -8.930 & 11.683 \vspace{0.2 cm} \\ 
Zr$_3$O & -7.560 & 19.146 & -6.184 & -7.508 & 17.676 \vspace{0.2 cm} \\ 
ZrSiO & -7.196 & 14.531 & -6.145 & -7.164 & 13.600 \\ 
& \\
\hline
\end{tabular}\\
\label{table:enthalpy-Si}
\end{table}

\begin{table}[!ht]
\caption{Enthalpies (H) of all the considered phases under pressures of 6 GPa for In-Ge-Zr-O system. H$^o$ and V$_0$ is the total energy and equilibrium volume of the relaxed structure without external pressure.  U and V are the total energy and volume under the given pressure (P) and H=U+PV. 1 GPa $\approx$ 0.0062415 eV/\AA$^3$ was used.}
\begin{tabular}{cccccc}
\hline
& \\
material & H$_0$ (eV/atom) & V$_0$ (\AA$^{3}$/atom) & H (eV/atom) & U (eV/atom) & V (\AA$^{3}$/atom) \\
& \\
\hline 
& \\
Ge & -4.621 & 23.908 & -3.765 & -4.588 & 21.997  \vspace{0.2 cm} \\ 
In & -2.726 & 27.62 & -1.759 & -2.672 & 24.387 \vspace{0.2 cm} \\ 
Zr & -6.598 & 25.235 & -5.682 & -6.573 & 23.796 \vspace{0.2 cm} \\
GeO$_2$ & -6.47 & 9.774 & -6.109 & -6.465 & 9.499 \vspace{0.2 cm} \\ 
GeZrO$_4$ & -7.748 & 11.034 & -7.341 & -7.743 & 10.725 \vspace{0.2 cm} \\ 
P-In$_2$Ge$_2$O$_7$ & -6.112 & 10.987 & -5.706 & -6.106 & 10.668 \vspace{0.2 cm} \\ 
T-In$_2$Ge$_2$O$_7$ & -6.127 & 13.519 & -5.632 & -6.116 & 12.945 \vspace{0.2 cm} \\ 
In$_2$O$_3$ & -5.676 & 13.683 & -5.173 & -5.666 & 13.170 \vspace{0.2 cm} \\
ZrGe & -6.375 & 19.763 & -5.652 & -6.359 & 18.867 \vspace{0.2 cm} \\ 
ZrO$_2$ & -8.964 & 12.586 & -8.501 & -8.955 & 12.126 \vspace{0.2 cm} \\ 
Zr$_3$O & -7.56 & 19.146 & -6.858 & -7.545 & 18.339 \\ 
& \\
\hline
\end{tabular}\\
\label{table:enthalpy-Ge}
\end{table}

\begin{table}
\caption{Four-phase equilibrium regions of T-In$_2$Si$_2$O$_7$ in the quaternary In-Si-Zr-O chemical space. $\Delta \mu_{i}$ ($i$ = In, Si, Zr, O) are the deviations in the elemental chemical potential from the reference values. -30 meV/atom is added to total energy of In$_2$Si$_2$O$_7$ to account for the limitation of the FERE approach.}

\begin{tabular}{cccccl}
\hline
&&&&\\
Vertex \  & $\Delta \mu_{\mathrm{In}}$ \ & $\Delta \mu_{\mathrm{Si}}$ \ & $\Delta \mu_{\mathrm{Zr}}$ \ & $\Delta \mu_{\mathrm{O}}$ \ & Phases in equilibrium with In$_2$Si$_2$O$_7$\\
& (eV) & (eV) & (eV)& (eV)&\\
&&&&&\\
\hline
&&&&&\\
V1& 0.000 &  -2.728 & -5.196 & -3.246 &  In, SiO$_2$, ZrSiO$_4$, In$_2$Si$_2$O$_7$  \\
V2& 0.000 &  -3.035 & -5.240 & -3.158 &  In, In$_2$O$_3$, ZrSiO$_4$, In$_2$Si$_2$O$_7$  \\
V3& -4.737 &  -9.351 & -11.556 & 0.000 &  O$_2$, SiO$_2$, ZrSiO$_4$, In$_2$Si$_2$O$_7$  \\
V4& -4.869 &  -9.220 & -11.688 & 0.000 &  O$_2$, In$_2$O$_3$, ZrSiO$_4$, In$_2$Si$_2$O$_7$  \\
&&&&&\\
\hline
\end{tabular}\\
\label{table:stab-Si-1}
\end{table}

\begin{table}
\caption{Four-phase equilibrium regions of P-In$_2$Si$_2$O$_7$ under 12 GPa in the quaternary In-Si-Zr-O chemical space. $\Delta \mu_{i}$ ($i$ = In, Si, Zr, O) are the deviations in the elemental chemical potential from the reference values at 12 GPa. H(P=12 GPa) were used instead of H$^{0}$.}

\begin{tabular}{cccccl}
\hline
&&&&\\
Vertex \  & $\Delta \mu_{\mathrm{In}}$ \ & $\Delta \mu_{\mathrm{Si}}$ \ & $\Delta \mu_{\mathrm{Zr}}$ \ & $\Delta \mu_{\mathrm{O}}$ \ & Phases in equilibrium with In$_2$Si$_2$O$_7$\\
& (eV) & (eV) & (eV)& (eV)&\\
&&&&&\\
\hline
&&&&&\\
V1& 0.000 &  -2.728 & -5.073 & -2.836 &  In, r-SiO$_2$, ZrSiO$_4$, In$_2$Si$_2$O$_7$  \\
V2& 0.000 &  -3.854 & -5.120 & -2.741 &  In, In$_2$O$_3$, ZrSiO$_4$, In$_2$Si$_2$O$_7$  \\
V3& -4.112 &  -9.337 & -10.603 & 0.000 &  O$_2$, In$_2$O$_3$, ZrSiO$_4$, In$_2$Si$_2$O$_7$  \\
V4& -4.254 &  -9.195 & -10.744 & 0.000 &  O$_2$, r-SiO$_2$, ZrSiO$_4$, In$_2$Si$_2$O$_7$  \\
&&&&&\\
\hline
\end{tabular}\\
\label{table:stab-Si-2}
\end{table}

\begin{table}
\caption{Four-phase equilibrium regions of T-In$_2$Ge$_2$O$_7$ in the quaternary In-Ge-Zr-O chemical space. $\Delta \mu_{i}$ ($i$ = In, Ge, Zr, O) are the deviations in the elemental chemical potential from the reference values.}

\begin{tabular}{cccccl}
\hline
&&&&\\
Vertex \  & $\Delta \mu_{\mathrm{In}}$ \ & $\Delta \mu_{\mathrm{Ge}}$ \ & $\Delta \mu_{\mathrm{Zr}}$ \ & $\Delta \mu_{\mathrm{O}}$ \ & Phases in equilibrium with In$_2$Ge$_2$O$_7$\\
& (eV) & (eV) & (eV)& (eV)&\\
&&&&&\\
\hline
&&&&&\\
V1& -0.797 &  -0.597 & -6.333 & -2.627 &  InGeO$_3$, In$_2$O$_3$, ZrGeO$_4$, In$_2$Ge$_2$O$_7$  \\
V2& -0.898 &  -0.496 & -6.434 & -2.627 &  InGeO$_3$, GeO$_2$, ZrGeO$_4$, In$_2$Ge$_2$O$_7$  \\
V3& -4.738 &  -5.851 & -11.587 & 0.000 &  O$_2$, GeO$_2$, ZrGeO$_4$, In$_2$Ge$_2$O$_7$  \\
V4& -4.838 &  -5.750 & -11.688 & 0.000 &  O$_2$, In$_2$O$_3$, ZrGeO$_4$, In$_2$Ge$_2$O$_7$  \\
&&&&&\\
\hline
\end{tabular}\\
\label{table:stab-Ge-1}
\end{table}

\begin{table}
\caption{Four-phase equilibrium regions of P-In$_2$Ge$_2$O$_7$ under 6 GPa in the quaternary In-Ge-Zr-O chemical space. $\Delta \mu_{i}$ ($i$ = In, Ge, Zr, O) are the deviations in the elemental chemical potential from the reference values at 6 GPa. H(P=6 GPa) were used instead of H$^{0}$}

\begin{tabular}{cccccl}
\hline
&&&&\\
Vertex \  & $\Delta \mu_{\mathrm{In}}$ \ & $\Delta \mu_{\mathrm{Ge}}$ \ & $\Delta \mu_{\mathrm{Zr}}$ \ & $\Delta \mu_{\mathrm{O}}$ \ & Phases in equilibrium with In$_2$Ge$_2$O$_7$\\
& (eV) & (eV) & (eV)& (eV)&\\
&&&&&\\
\hline
&&&&&\\
V1& -0.215 &  0.000 & -5.475 & -2.823 &  Ge, In$_2$O$_3$, ZrGeO$_4$, In$_2$Ge$_2$O$_7$  \\
V2& -0.431 &  0.000 & -5.722 & -2.761 &  Ge, GeO$_2$, ZrGeO$_4$, In$_2$Ge$_2$O$_7$  \\
V3& -4.573 &  -5.523 & -11.245 & 0.000 &  O$_2$, GeO$_2$, ZrGeO$_4$, In$_2$Ge$_2$O$_7$  \\
V4& -4.450 &  -5.646 & -11.121 & 0.000 &  O$_2$, In$_2$O$_3$, ZrGeO$_4$, In$_2$Ge$_2$O$_7$  \\
&&&&&\\
\hline
\end{tabular}\\
\label{table:stab-Ge-2}
\end{table}

\begin{table}[!ht]

\caption {Predicted material properties of In$_{2}$X$_{2}$O$_{7}$ (X=Si, Ge) using PBE functional. SG is the spacegroup. $a$, $b$, $c$, and/or $\beta$ are lattice parameters relevant to pyrochlore (p) and thorvetite (t) structures. The relaxed structures were standardized by \textit{spglib}. Superscript of $GW$ indicates PBE band gap ($E_g$) corrected by the GW method. m$^*_b$  and $N_b$ are average band effective mass and band degeneracy, respectively, for electron (e) and hole (h). $\epsilon_{ion}$ and $\epsilon_{el}$ are the isotopically averaged  ionic and electronic contributions, respectively, to static dielectric constant.}\label{tab:parameter} 
\begin{ruledtabular}
    \scriptsize
    \begin{tabular}{llllllllllll}
        material & SG & $a$ (\AA) & $b$ (\AA) & $c$ (\AA) & $\beta$ ($^\circ$) & volume (\AA$^3$) & $E_g^{PBE}$ (eV) & $E_g^{GW}$ (eV) & m$^*_{b,e}$ / m$^*_{e,h}$ ($m_e$)  & $N_{b,e}$/$N_{b,h}$  & $\epsilon_{el}$/$\epsilon_{ion}$ \vspace{0.2 cm} \\ 
        T-In$_2$Ge$_2$O$_7$ & 12 & 6.745 & 8.965 & 5.005 & 102.37 & 295.63 & 1.82 & 4.07 & 0.22/5.83 & 1/1 & 3.92/10.62 \vspace{0.2 cm} \\ 
        P-In$_2$Ge$_2$O$_7$ & 227 & 9.876 & - & - & - & 963.35 & 1.10 & 2.77 & 0.29/4.28 & 1/5 & 4.83/16.97 \vspace{0.2 cm} \\ 
        T-In$_2$Si$_2$O$_7$ & 12  & 6.709 & 8.738 & 4.77 & 102.72 & 272.75 & 2.73 & 5.45 & 0.19/8.19 & 1/2 & 3.43/5.71 \vspace{0.2 cm} \\ 
        P-In$_2$Si$_2$O$_7$ & 227  & 9.542 & - & - & - & 868.76 & 1.83 & 3.90 & 0.35/7.24 & 1/2 & 4.21/15.45\\ 
    \end{tabular}
\end{ruledtabular}
\end{table}

\begin{table}[!htb]
\caption {Defect formation energy with Fermi energy at valence band maximum, $\Delta E_{\mathrm{D}}(E_{\mathrm{F}}=E_v)$, for donor impurity candidates in P-In$_{2}$Ge$_{2}$O$_{2}$ under O-poor condition.}\label{tab:DFE_candidates} 
\begin{ruledtabular}
\begin{tabular}{llllllll}
Donor impurity                & Zr$\mathrm{_{In}^{+1}}$ &  Sn$\mathrm{_{In}^{+1}}$ &  Hf$\mathrm{_{In}^{+1}}$ &   Pb$\mathrm{_{In}^{+1}}$ &  As$\mathrm{_{Ge}^{+1}}$ & V$\mathrm{_{Ge}^{+1}}$ & Bi$\mathrm{_{Ge}^{+1}}$ \vspace{0.2 cm} \\ 
$\Delta E_{\mathrm{D}}(E_{\mathrm{F}}=E_v)$ &   -2.389  & -1.895 & -2.326 & 0.249 & -2.101 & -0.999 & 1.033  \\
\end{tabular}
\end{ruledtabular}
\end{table}


\bibliography{In2Ge2O7}